\documentclass[12pt,halfline,a4paper]{ouparticle}
\graphicspath{{Figure/}}
\usepackage{multirow}
\usepackage{multicol}
\usepackage{amssymb}
\usepackage{amsmath}
\usepackage{ntheorem}
\usepackage{amsbsy}
\usepackage[round]{natbib}   
\usepackage{bbold}
\usepackage{authblk}

\newtheorem{theorem}{Theorem}

\newtheorem{assumption}{Assumption}

\newtheorem{lemma}{Lemma}
\newtheorem{remark}{Remark}
 \def\bx{{\mathbf x}}
 \def \bb{{\pmb{\beta}}}
 \def\b1{{\mathbf 1}}
 
\def\bD{{\mathbf D}}

\def\be{{\mathbf e}}

\def\bone{{\mathbf{1}}}
\def\bz{{\pmb{z}}} \def \bnu{{ \pmb{\nu}}}
\def \bphi{{\pmb \phi}}

\def\RR{\mathbb{R}} 

\def \mL{ \mathcal{L}}
\def \bo{{\mathcal{O}}}
\def \b0{{ \mathbf 0}}
\def \blp{\left (}
\def \brp{\right )}
\def\bSigma{{\mathbf \Sigma}}

\newcommand{\argmin}{\operatornamewithlimits{argmin}}
\newcommand{\indep}{\rotatebox[origin=c]{90}{$\models$}}
\newcommand*{\QEDB}{\hfill\ensuremath{\square}}%
\newtheorem{example}{Example}[section]

\date{}
\begin{document}

\title{A High-dimensional M-estimator Framework for Bi-level Variable Selection}

\author[1]{Bin Luo \thanks{bin.luo2@duke.edu} }
\author[2]{Xiaoli Gao}
\affil[1]{Department of Biostatistics and Bioinformatics, Duke University}
\affil[2]{Department of Mathematics and Statistics, The University of North Carolina at Greensboro}

\abstract{
In high-dimensional data analysis, bi-level sparsity is often assumed when covariates function group-wisely and sparsity can appear either at the group level or within certain groups. In such cases, an ideal model should be able to encourage the bi-level variable selection consistently. Bi-level variable selection has become even more challenging when data have heavy-tailed distribution or outliers exist in random errors and covariates. In this paper, we study a framework of high-dimensional M-estimation  for bi-level variable selection. This framework encourages bi-level sparsity through a computationally efficient two-stage procedure. In theory, we provide sufficient conditions under which our two-stage penalized M-estimator possesses simultaneous local estimation consistency  and the bi-level variable selection consistency if certain nonconvex  penalty functions are used at the group level. Both our simulation studies and real data analysis demonstrate satisfactory finite sample performance of the proposed estimators under different irregular settings.}


\keywords{Bi-level variable selection,  Estimation consistency, High dimensionality, M-estimation, Non-convexity.
}

\maketitle

\section{Introduction}

Covariates often function group-wisely in many applications. For example, in gene expression analysis, genes from the same biological pathways may exhibit similar activities. In high-dimensional linear regression, penalized least squares approaches with penalties incorporating grouping structures have become very popular in recent decades. \cite{yuan2006model} proposed the group Lasso, as a natural extension of the Lasso \citep{tibshirani1996regression}, to select variables at the group level by applying the Lasso penalty on the $\ell_2$ norm of coefficients associated with each group of variables  in penalized least squares regression (LS-GLasso). To address the bias and inconsistency of the group Lasso estimator in high-dimensional settings,  several methods have been investigated, including the adaptive  group Lasso \citep{wei2010consistent}, the $\ell_2$-norm MCP \citep{huang2012selective}, the $\ell_2$-norm SCAD \citep{guo2015model}, among others. However, the above approaches encourage only ``all-in" or ``all-out" variable selection at the group level. To further encourage the sparsity within certain groups, extensive methods have been proposed to perform bi-level variable selection. See for example the group Bridge \citep{huang2009group}, the sparse group Lasso \citep{friedman2010note, simon2013sparse}, the concave $\ell_1$-norm group penalty \citep{jiang2014concave}, the composite MCP \citep{breheny2009penalized}, the group exponential lasso \citep{breheny2015group}, among others. See \cite{huang2012selective} for a complete review.

When the data dimensionality grows much faster than the sample size, irregular settings often appear, such as the response and a large number of variables are contaminated or heavy-tailed. It has been shown that the LS-GLasso is estimation consistent when the random errors are sub-Gaussian \citep{wei2010consistent}. However, the quadratic loss in LS-GLasso is non-robust to outliers and the estimator is no longer consistent if the random errors wildly deviate from sub-Gaussian distribution. In addition, the required restricted eigenvalue condition on the design matrix may not hold if the predictors are non-Gaussian. 

To tackle the problem of heavy-tailed random errors in high-dimensional settings, a few robust penalized approaches have been recently studied. \cite{lilly2015robust} proposed the penalized least absolute deviation (LAD) estimator with the group Lasso penalty to relieve the model's sensitivity due to the existence of outliers in random errors. This method was also extended to the weighted LAD group Lasso when some predictors are contaminated or heavy-tailed. \cite{wang2016robust} investigated a general penalized M-estimators framework using convex loss functions and concave $\ell_2$-norm penalties for the partially linear model with grouped covariates. However, those robust methods can only select variables at the group level and thus do not perform bi-level variable selection. In the example of gene expression study, while the data may be heavy-tailed or contaminated due to the complex data generation procedures, we may be still interested in selecting important genes as well as important groups.



Additionally, the above robust methods require the loss function to be convex. It is well known that the convex loss functions such as Huber loss and LAD loss do not downweight the very large residuals due to their convexity.  \cite{shevlyakov2008redescending} showed that re-descending M-estimators with non-convex loss function possess certain optimal robustness properties. In fact, there still lacks a systematic study of high-dimensional M-estimators that perform robust bi-level variable selection, allowing both loss and group penalty functions to be non-convex.

In this paper, we consider high-dimensional linear regression with grouped covariates, in irregular settings that the data (random errors and/or  covariates) may be contaminated or  heavy-tailed. In particular, we propose a novel high-dimensional bi-level variable selection method through a two-stage penalized M-estimator framework: penalized M-estimation with a concave $\ell_2$-norm penalty achieving the consistent group selection at the first stage, and a post-hard-thresholding operator to achieve the within-group sparsity at the second stage. Our perspective at the first stage is different from \cite{wang2016robust} since we allow the loss function to be non-convex and thus it is more general. In addition, our proposed two-stage framework is able to separate the group selection and the individual variable selection efficiently, since the post-hard-thresholding operator at the second stage nearly poses no additional computational  burden. More importantly, our framework includes a wide range of M-estimators with strong robustness if a redescending loss function is adopted.
Furthermore, we extend our framework to a more general setting by relaxing the sub-Gaussian assumption enforced on covariates. 

Theoretically, we investigate the statistical properties of our proposed two-stage framework with  weak assumptions on both random errors and covariates. We first show that with certain mild conditions on the loss function, a penalized M-estimator at the first stage has the local estimation consistency at the minimax rate enjoyed by the LS-GLasso. We further establish that with an appropriate group concave $\ell_2$-norm penalty,  the estimator from our first stage has a group-level oracle property. We then show that these nice statistical properties can be carried over directly to the post-hard-thresholding estimators at the second stage and thus we establish its bi-level variable selection consistency. The theoretical results allow the dimensionality of data to grow with the sample size at an almost exponential rate.

Computationally, we propose to implement an efficient algorithm through a two-step optimization procedure. We compare the performance of estimators generated from different types of loss functions (e.g. the Huber loss and Cauchy loss) combined with a concave penalty (e.g. MCP penalty). Our numerical results demonstrate satisfactory finite sample performances of the proposed estimators under different settings. Additionally, it also suggests that a well-behaved two-stage M-estimator can be usually obtained by considering a re-descending loss (e.g. Cauchy loss) with a concave penalty, when the data are heavy-tailed or strongly contaminated.

The remainder of our paper is organized as follows. In Section 2, we introduce a basic setup for our two-stage penalized M-estimator framework. In Section 3, we present the statistical properties of our proposed bi-level M-estimators under some sufficient conditions. We discuss the implementation of the two-stage M-estimators in Section 4. 
In Section 5, we conduct some simulation studies to demonstrate the performance of the proposed estimators under different settings. We also apply the proposed estimators for NCI-60 data analysis and illustrate all results in Section 6. 
Section 7 concludes and summarizes the paper. 
All technical proofs are relegated to Appendix.


\section{The Two-stage M-estimator Framework} \label{sec:2}
Let's consider a high-dimensional data with $p$ covariates from $J$ non-overlapping groups. A linear regression model can be written as
\begin{equation} \label{linear model}
   	 y_i=\sum_{j=1}^{J}\bx_{ij}^T \bb_j^*+\epsilon_i, \quad i=1,\cdots,n,
\end{equation}
where  $\epsilon_i$s
 are independent and identically distributed (i.i.d) random errors,
 $\bx_{ij}$s are i.i.d $d_j$-dimensional covariate vectors corresponding to the $j$th group,   $\bb_j^*$ is the $d_j$-dimensional true regression coefficient  vector of the $j$th group. Then $p= \sum_{j=1}^{J} d_j$. Let $\bx_i=(\bx_{i1}^T, \cdots, \bx_{iJ}^T)^T$ and $\bb^* = (\bb_1^{*T}, \cdots, \bb_J^{*T})^T$.  We assume the independence between covariates $\bx_{i}$ and random errors $\epsilon_i$ for the sake of simplicity. We also assume the group sparsity condition of the model: there exists  $S \subseteq \{1, \cdots, J\}$ such that $\bb_j^* = \bf 0$ for all $j \notin S$. Note that we allow the within-group sparsity on some $\bb^*_j \ne \b0$ and thus there exists bi-level sparsity on the coefficient vector $\bb^*$.

\paragraph{Some More Notations} We use bold symbols to denote matrices or vectors.  Let $\beta_m$ be the $m$th element of $\bb \in \RR^p$. For any $A \subseteq \{1,2, \cdots, p\}$, we denote $\bb_A=(\beta_m, ~m \in A)^T$ a coefficient sub-vector with indexes in $A$. Define $d_a := \max_{1 \le j \le J}d_j$, $d_b := \min_{1 \le j \le J}d_j$, $d:=\sqrt{\frac{d_a}{d_b}}$.  Let $I_j \subseteq \{1,2, \cdots, p\}$ denote the index set of coefficients in group $j$. Then $I_S:=\bigcup_{j \in S} I_j$ includes all indexes of coefficients in those important groups. Let $I_0=\{m:\beta_m^*\ne 0, 1 \le m \le p\}$ and thus $I_0 \subseteq I_S$. 
Define  $\bb^{*G}_{\min}:= \min_{j \in S} \|\bb^*_j\|_2$ 
as the minimum group strength on $\bb^*$, where $\|\cdot\|_2$ is the $\ell_2$ norm.
Define 
 $\bb^{*I}_{\min}:= \min_{m \in I_0} |\beta_m^*|$ as the  minimum individual signal strength on $\bb^*$. 
Let $s=|S|$ and $k = |I_S|$ be the number of  important groups and number of variables among all important groups, respectively.  
 We denote $u_+=\max(u,0)$ for any $u \in \RR$.

\paragraph{Our Proposed M-estimator Framework for Bi-level Variable Selection} 
To perform an efficient bi-level variable selection with robustness
for the existence of possible data contamination or heavy-tailed distribution between $\epsilon_i$ and $\bx_i$,
we propose the following two-stage penalized M-estimator framework: 
\begin{itemize}
    \item {\it Group Penalization (GP) Stage.} First we perform penalized M-estimation with a group concave penalty achieving the between-group sparsity:
   $$
        \hat{\bb} \in \argmin_{ \bb \in \RR^{p}, \|\bb\|_1\le R} \left\{ \mL_n(\bb)+\sum_{j=1}^{J} \rho(\|\bb_j\|_2,\sqrt{d_j}\lambda ) \right\}.
        $$
    \item{\it Hard-thresholding (HT) Stage.} Then we apply a post-hard-thresholding operator on $\hat{\bb}$:
    \begin{equation} \label{eq:final-estimator}
        \hat{\bb}^h (\theta)=  \hat{\bb}\cdot I(|\hat{\bb}|\ge \theta)
    \end{equation}
    where ``$\cdot$'' and ``$\ge$'' in \eqref{eq:final-estimator}
    are elementary-wise. 
\end{itemize}
Here $\mL_n$ is an empirical loss function that may produce a robust solution and $\rho$ is a penalty function, which encourages the group sparsity in the solution. Note that $\lambda$ and $\theta$ are tuning parameters controlling the between-group and within-group sparsity, respectively.  We include the side condition $\|\bb\|_1 \le R$ for $R \ge \|\bb^*\|_1$ in the  Group Penalization Stage in order to guarantee the existence of local/global optima, for the case where the loss or regularizer may be non-convex. In real applications, $R$ can be a sufficiently large number.

Let $l: \RR \mapsto \RR $ denote a residual function, or a loss function, defined on each observation pair ($\bx_i, y_i$). Then the above Group Penalization Stage becomes
 \begin{equation} \label{eq:group-penalization}
        \hat{\bb} \in \argmin_{ \bb \in \RR^{p}, \|\bb\|_1\le R}
        \left\{ 
        \frac{1}{n} \sum_{i=1}^{n}l(y_i - \bx_i^T\bb) 
        +\sum_{j=1}^{J} \rho(\|\bb_j\|_2,\sqrt{d_j}\lambda ) 
        \right\}.
    \end{equation}
With a well chosen $l$, the penalized M-estimator from (\ref{eq:group-penalization}) can be robust to heavy-tailed random error $\epsilon_i$. Some typical robust loss functions $l$ include:
\begin{itemize}
    \item 
{\bf Huber Loss}
\begin{equation*}
l(u)=\begin{cases}
\frac{u^2}{2}  & \text{if } |u| \le \alpha, \\
\alpha |u| - \frac{\alpha^2}{2} & \text{if } |u| \ge \alpha.
\end{cases} 
\end{equation*}
\item {\bf Tukey's biweight Loss}
\begin{equation*}
l(u)=\begin{cases}
\frac{\alpha^2}{6}(1-(1-\frac{u^2}{\alpha^2})^3) & \text{if } |u| \le \alpha, \\
\frac{\alpha^2}{6} & \text{if } |u| \ge \alpha.
\end{cases}
\end{equation*}
\item {\bf Cauchy Loss}
\begin{equation*}
l(u)=\frac{\alpha^2}{2}\log\left (1+\frac{u^2}{\alpha^2} \right).
\end{equation*}
\end{itemize}

The derivatives of the above three loss functions are bounded and thus they can mitigate the effect of larger residuals. In particular, the Tukey's biweight loss and Cauchy loss produce re-descending $M$-estimators. From the robust regression literature, we call an $M$-estimator re-descending if there exists $u_0>0$ such that $|l'(u)|=0$ or decrease to 0 smoothly, for all $|u| \ge u_0$. In that case, strong robustness is obtained by ignoring the large outliers completely. See more discussions in \cite{muller2004redescending} and   \cite{shevlyakov2008redescending}.

Whereas the robust loss function in \eqref{eq:group-penalization} takes into account the contamination or heavy-tailed distribution in error $\epsilon_i$, a single outlier in $\bx_i$ may still cause the corresponding estimator to perform arbitrarily badly. To downweight large values of $\bx_i$, we extend the  Group Penalization Stage in \eqref{eq:group-penalization} to   \begin{equation} \label{eq:ex-group-penalization}
        \hat{\bb} \in \argmin_{ \bb \in \RR^{p}, \|\bb\|_1\le R} \left\{  \frac{1}{n} \sum_{i=1}^{n} \frac{w(\bx_i)}{v(\bx_i)}l((y_i-\bx_i^T \bb)v(\bx_i))+\sum_{j=1}^{J} \rho(\|\bb_j\|_2,\sqrt{d_j}\lambda ) \right\},
    \end{equation}
where $w,v$ are weight functions such that $w,v>0$. A few options for choosing those weight functions can be found in \cite{mallows1975some}, \cite{hill1977robust}, \cite{merrill1971bad} and \cite{loh2017statistical}. 

Since $\bb_j^* = \b0$ for $j \notin S$, we need the Group Penalization Stage to generate sparse solutions between groups. 
In particular, we require the penalty function $\rho$ in \eqref{eq:ex-group-penalization} to  satisfy  amendable  properties listed in Assumption 1.
\begin{assumption}[Penalty Function Assumptions] \label{as:penalty}
	$\rho: \RR\times \RR \mapsto \RR$ is a scalar function that satisfies the following conditions: 
\begin{itemize}
    \item[(i)] For any fixed $t \in \RR^+$,  the function $\lambda \mapsto \rho(t,\lambda)$  is non-decreasing on $\RR^+$.
    \item[(ii)] There exists a scalar function $g:\RR^+ \mapsto \RR^+$ such that for any $r \in [1, \infty)$, $\frac{\rho(t,r\lambda )}{\rho(t, \lambda)} \le g(r)$ for all $t,\lambda \in \RR^+$.
\end{itemize}
\begin{itemize}
\item[(iii)] The function $t \mapsto \rho(t, \lambda)$ is symmetric around zero and $\rho(0,\lambda)=0$, given any fixed $\lambda \in \RR$.
\item[(iv)] The function $t \mapsto \rho(t,\lambda)$  is non-decreasing on $\RR^+$, given any fixed $\lambda \in \RR$.
\item[(v)] The function $t \mapsto \frac{\rho(t,\lambda)}{t}$ is non-increasing on $\RR^+$, given any fixed $\lambda \in \RR$.
\item[(vi)] The function $t \mapsto \rho(t,\lambda)$  is differentiable for $t \ne 0$, given any fixed $\lambda \in \RR$.
\item[(vii)] $ \lim_{t \to 0^+} \frac{\partial \rho(t,\lambda)}{\partial t}= \lambda$, given any fixed $\lambda \in \RR$.
\item[(viii)] There exists $\mu > 0$ such that the function $t \mapsto \rho(t,\lambda) + \frac{\mu}{2}t^2$is convex, given any fixed $\lambda \in \RR$.
\item[(ix)] There exists $\delta \in (0, \infty)$ such that $\frac{\partial \rho(t,\lambda)}{\partial t}=0$ for all $t \ge \delta \lambda$, given any fixed $\lambda \in \RR$. 
\end{itemize} 

\end{assumption}
The properties (iii-ix) in Assumption 1 are related to the penalty functions  studied in \citet{loh2017statistical} and  \citet{loh2015regularized}. Adopting the notation from \citet{loh2017statistical}, we consider $\rho$ to be $\mu$-amenable if $\rho$ satisfies conditions (i)-(viii). If $\rho$ also satisfies condition (ix), we say that $\rho$ is $(\mu, \delta)$-amenable. 
In general, if $\rho$ is $\mu$-amenable, $ q(t,\lambda):=\lambda |t|-\rho(t,\lambda)$ is partially differentiable for $t\ne0$. However, properties (iii) and (vii) imply that $ \lim_{t \to 0} \frac{\partial q(t,\lambda)}{\partial t}= 0$. Therefore, we can define $ \frac{\partial q(t,\lambda)}{\partial t} \left. \right|_{t=0}= 0$ and have $q(t,\lambda)$ everywhere differentiable with respect to $t$.
Define the vector version $q_\lambda(\bb):=\sum_{j=1}^{J} q(\|\bb_j\|_2,\sqrt{d_j}\lambda)$ accordingly. It is easy to see that there exists $\mu>0$ such that $\frac{\mu}{2}\|\bb\|_2^2 - q_\lambda(\bb)$ is convex. This property is important for both computational implementation and theoretical investigation of the group selection properties.

Some popular choices of amenable penalty functions include Lasso \citep{tibshirani1996regression}, SCAD \citep{fan2001variable}, and MCP  \citep{zhang2010nearly}. Below are the expressions of Lasso and MCP, which will be considered in our numerical analyses.
\begin{itemize}
\item The {\bf Lasso} penalty $\rho(t, \lambda)=\lambda |t|$ is $0$-amenable but not $(0,\delta)$-amenable for any $\delta <\infty$.
\item The {\bf MCP} penalty takes the form 
\begin{equation*}
\rho(t,\lambda)={\rm sign}(t) \lambda \int_{0}^{|t|}\left(1-\frac{z}{\lambda b}\right)_+ dz,
\end{equation*}
where $b>0$ is fixed. The MCP penalty is $(\mu, \delta)$-amenable with $\mu=\frac{1}{b}$ and $\delta=b$.   
\end{itemize}
It has been shown that a folded concave penalty, such as the SCAD or MCP, often has better variable selection properties than the convex penalty including the Lasso.

\section{Statistical Properties}
In this section, we present our theoretical results for the proposed two-stage penalized M-estimator framework. We begin with statistical properties of the estimator $\hat{\bb}$ in program (\ref{eq:ex-group-penalization}) generated from the Group Penalization Stage. On the one hand, we show a general non-asymptotic bound of the estimation error and establish the local estimation consistency of $\hat{\bb}$ at the minimax rate enjoyed by the LS-GLasso, under certain mild conditions. On the other hand, 
we  show that the estimator $\hat{\bb}$ in fact equals the local oracle solution with the correct group support and thus obtain the group-level oracle properties. Finally, we show that those nice statistical properties of $\hat{\bb}$ can be carried over to the hard-thresholding stage and thus we establish the bi-level variable selection consistency
of $\hat{\bb}^h$. All proofs are given in Appendix. 
 
As introduced in  (\ref{eq:ex-group-penalization}), the loss function in the two-stage penalized M-estimator framework takes the following form,
\begin{equation} \label{eq:gloss}
\mL_{n}(\bb)=\frac{1}{n} \sum_{i=1}^{n} \frac{w(\bx_i)}{v(\bx_i)}l((y_i-\bx_i^T \bb)v(\bx_i)).
\end{equation}
To obtain the estimation consistency, we make the following assumptions on the residual function $l$.
\begin{assumption}[Loss Function Assumptions] \label{as:loss}
$l: \RR \mapsto \RR$ is a  scalar function with the existence of the first derivative $l'$ everywhere and the second derivative $l''$ almost everywhere. In addition,
	\begin{enumerate} 
		\item[(i)] there exists a constant $0 < k_1 < \infty$ such that $|l'(u)| \le k_1$ for all $ u \in \RR$.
		\item[(ii)]$l'$ is Lipschitz such that $|l'(x) - l'(y)| \le k_2|x-y|$, for all $x, y \in \RR$ and some $0 <k_2 <\infty$.
	\end{enumerate}
\end{assumption}

Note that Assumption \ref{as:loss}(i) requires bounded derivative of the loss function, which can limit the effect of large residuals and thus achieve certain robustness. Assumption \ref{as:loss}(ii) indicates that $|l''(u)| < k_2$ for all $u\in \RR$ where $l''(u)$ exists. The above assumptions actually cover a wide range of loss functions, including Huber loss, Hampel loss,  Tukey's biweight and Cauchy loss. 

We now make some assumptions on both random error $\epsilon$ and covariate vector $\bx$.

\begin{assumption}[Error and Covariate Assumptions] \label{as:x} 
For $w(\bx)$ and $v(\bx)$ given in \eqref{eq:gloss}, the random error $\epsilon$ with $E[\epsilon]=0$ and covariate vector $\bx$ with $E[\bx]=\b0$ satisfy:
	\begin{enumerate}
		\item[(i)] for any $\pmb \nu \in \RR^p$, $w(\bx)\bx^T\pmb \nu$ is sub-Gaussian with parameter at most $k_0^2\|\pmb \nu\|_2^2$.
		\item[(ii)] either
		(a) $v(\bx)=1$ and $E[w(\bx)\bx]=\b0$, or (b) $E[l'( v(\bx)\epsilon)|\bx]=0$.
	\end{enumerate}
\end{assumption}

Note that Assumption \ref{as:x}(i) and (ii)(a) hold when $\bx_i^T\pmb \nu$ is sub-Gaussian for any $\pmb \nu \in \RR^p$ and $w(\bx)=1$. If covariate $\bx$ is contaminated or heavy-tailed, Assumption \ref{as:x}(i) nonetheless holds with some proper choices of $w(\bx)$ (e.g. $w(\bx)\bx^T \pmb \nu$ is bounded for any $\pmb \nu \in \RR$), which potentially relaxes the sub-Gaussian assumption on $\bx$. Assumption \ref{as:x}(ii)(b) holds when $l'$ is an odd function and $\epsilon$ follows a symmetric distribution. Despite the possible mild condition of symmetry, the assumptions above are independent of the distribution of $\epsilon$, allowing the additive error $\epsilon$ to be heavy-tailed or contaminated.

In order to obtain the estimation  consistency for $\hat{\bb}$ in (\ref{eq:ex-group-penalization}), we also require the loss function $\mL_{n}$ to satisfy the following local Restricted Strong Convexity (RSC) condition. This RSC condition was also investigated in \cite{loh2015regularized} and \cite{loh2017statistical}.

\begin{assumption}[RSC condition]\label{as:RSC}
	 There exist $\gamma$, $\tau>0$ and a radius $r>0$ such that the loss function
	$\mL_{n}$ in  \eqref{eq:gloss} satisfies
	\begin{equation} \label{eq: RSC}
	\langle \nabla \mL_{n}(\bb_1)- \nabla \mL_{n}(\bb_2), \bb_1-\bb_2 \rangle \ge \gamma \|\bb_1-\bb_2\|_2^2 - \tau \frac{\log p}{n}\|\bb_1 - \bb_2\|_1^2,
	\end{equation}
	where $\bb_j \in \RR^p$ such that $\|\bb_j-\bb^*\|_2\le r$ for $j=1, 2$.
\end{assumption}

Note that the RSC assumption is only imposed
on $\mL_{n}$ inside the ball of radius $r$ centered at $\bb^*$. Thus the loss function used for robust regression can be wildly nonconvex while it is away from  the origin. The ball of radius $r$ essentially specifies a local region around $\bb^*$ in which stationary points of program (\ref{eq:ex-group-penalization}) are well-behaved. We call such a region  the RSC region.

We present the estimation consistency result concerning estimator $\hat{\bb}$ in the following Theorem \ref{Tm:1}.
\begin{theorem}\label{Tm:1} 
Suppose the random error and covariates satisfy Assumption \ref{as:x} and  $\mL_{n}$   in \eqref{eq:gloss} satisfies Assumption \ref{as:loss}. Then we have the following results.
\begin{enumerate}
    \item [(i)]	It holds with probability at least $1-2\exp(-C_2\log p)$ that $\mL_{n}$ satisfies
	\begin{equation} \label{eq:Ln-bound}
	\|\nabla \mL_{n}(\bb^*)\|_{\infty} \le C_1\sqrt{\frac{\log p}{n}}.
	\end{equation}
	\item[(ii)] 
		Suppose $\mL_n$ satisfies the RSC condition in Assumption \ref{as:RSC} with $\bb_2=\bb^*$ and $\rho$ is $\mu$-amenable with $\frac{3}{4}\mu < \gamma$ in Assumption \ref{as:penalty}. Let $\hat{\bb}$ be a local estimator in (\ref{eq:ex-group-penalization})  in  the RSC region. Then for $n \ge Cr^{-2}d_as\log p$ and $\lambda \ge \max\{ 4 \|\nabla \mL_n(\bb^*)\|_\infty, 8\tau R \frac{\log p}{n}\}$,
	$\hat{\bb}$ exists and satisfies the bounds
	\begin{equation*}
	\|\hat{\bb}-\bb^*\|_2 \le \frac{6\sqrt{d_a}\lambda\sqrt{s}}{4\gamma - 3\mu} \text{ and  }  \|\hat{\bb}-\bb^*\|_1 \le \frac{6(1+3g(d))d_a\lambda s}{4\gamma - 3\mu}.
	\end{equation*}
\end{enumerate}

\end{theorem}

The statistical consistency result of Theorem \ref{Tm:1} holds even when the random errors are heavy-tailed, and the regressors lack the sub-Gaussian assumption. Theorem \ref{Tm:1}(ii) essentially gives general deterministic bounds of the estimation error, provided that the loss function $\mL_n$ satisfies the RSC condition and  the penalty function $\rho$ is $\mu$-amenable. In particular, Theorem \ref{Tm:1} shows that with high probability one can choose $\lambda=\mathcal{O}\left (\sqrt{\frac{\log p}{n}} \right)$ such that $\|\hat{\bb}-\bb^*\|_2=\mathcal{O}_p \left (\sqrt{\frac{d_a s\log p}{n}} \right)$ and $\|\hat{\bb}-\bb^*\|_1=\mathcal{O}_p \left (g(d)d_as\sqrt{\frac{\log p}{n}} \right)$. Hence if $d_a$ is finite, the estimator $\hat{\bb}$ at the Group Penalization Stage is statistically consistent at the minimax rate enjoyed by the LS-GLasso under the sub-Gaussian assumption.

\begin{remark}
 Recall that  $\tilde{\bb}$ is a stationary point of the optimization in (\ref{eq:ex-group-penalization}) if 
\begin{equation*}
    \langle \nabla \mL_{n}(\tilde{\bb}) + \nabla \rho_{\lambda}(\tilde{\bb}), \bb - \tilde{\bb} \rangle \ge 0,
\end{equation*}
for all feasible $\bb$ in a neighbour of $\tilde{\bb}$, where $\rho_{\lambda}(\bb)=\sum_{j=1}^{J} \rho(\|\bb_j\|_2,\sqrt{d_j}\lambda )$. Note that stationary points include both the interior local maxima as well as all local and global minima. The proof of Theorem \ref{Tm:1} in Appendix reveals that the estimation consistency result also holds for the stationary points in program (\ref{eq:ex-group-penalization}). Hence Theorem \ref{Tm:1} guarantees that all stationary points within the ball of radius $r$ centered at $\bb^*$ have local statistical consistency at the minimax rate enjoyed by the LS-GLasso. To simplify the notation, $\hat{\bb}$ also denotes the stationary points of program (\ref{eq:ex-group-penalization}).
\end{remark}

Next we establish the group-level oracle properties of estimator $\hat{\bb}$ in (\ref{eq:ex-group-penalization}). 
Suppose $I_S$ is given in advance,we define the group-level local oracle estimator as
\begin{equation} \label{eq:oracle}
    \hat{\bb}_{I_S}^{\mathcal{O}}:= \argmin_{\bb \in \RR^{I_S}:\|\bb - \bb^*\|_2 \le r} \left\{ \mL_{ n} (\bb) 
        \right \}.
\end{equation}
Let $\hat{\bb}^\bo := (\hat{\bb}^\bo_{I_S}, \b0_{I_S^c})$. 
The next theorem shows that when the penalty $\rho$ is $(\mu, \delta)$-amenable and conditions in Theorem \ref{Tm:1} are satisfied, the stationary point from (\ref{eq:ex-group-penalization}) within the local neighborhood of $\bb^*$ is actually unique and agrees with the group oracle estimator in (\ref{eq:oracle}). 
\begin{theorem} \label{Tm:2}
Suppose the penalty $\rho$ is $(\mu, \delta)$-amenable and conditions in Theorem \ref{Tm:1} hold. Suppose in addition that $v(\bx)x_j$ is sub-Gaussian for all $j=1,\cdots,p$, $\|\bb^*\|_1 \le \frac{R}{2}$ for some  $R >\frac{12(1+3g(d))d_a\lambda s}{4\gamma - 3\mu}$ and $\bb^{*G}_{\min} \ge C_8\sqrt{\frac{d_a\log p}{n}}$. Let $\hat{\bb}$ be a stationary point of program (\ref{eq:ex-group-penalization}) in the RSC region. Then for $n \ge C_0 k \log p$, $k^2\log k =\bo (\log p)$ and $\lambda=C_6\sqrt{\frac{\log p}{n}}$,  $\hat{\bb}$ satisfies $ supp (\hat{\bb})\subseteq I_S $ and $\hat{\bb}_{I_S} = \hat{\bb}_{I_S}^{\mathcal{O}}$ with probability at least $1-C_7 \exp(-C_4 \log p/k^2)$.
\end{theorem}

Theorem \ref{Tm:2} guarantees that the Group Penalization Stage in our proposed framework can recover the true group support with high probability, when the condition of minimum group signal strength is satisfied. Two most common $(\mu, \delta)$-amenable penalties are SCAD and MCP, as introduced in Section \ref{sec:2}. 

It has been shown that the GP Stage can select important covariates groups and provides consistent estimation for parameter $\bb^*$. We are now ready to establish statistical properties of $\hat{\bb}^h$ after the HT stage in our proposed framework. 
 We reveal in the following theorem that when the condition of minimum individual signal strength is satisfied, the estimate of the zero elements and the non-zero elements of $\bb^*$ after the GP Stage can then be well separated. Hence, there exist some thresholds that are able to filter out those non-important covariates within the selected important groups, and thus the HT Stage can perform bi-level variable selection consistently.


\begin{theorem} \label{Tm:3}
 Suppose conditions of Theorem \ref{Tm:2} hold and in addition that $\bb^{*I}_{\min} \ge C_5\sqrt{\frac{\log p}{kn}} + \theta$ and $\theta > C_5\sqrt{\frac{\log p}{kn}} $. With probability at least $1-C_7 \exp(-C_4 \log p/k^2)$,  the hard-thresholding estimator $\hat{\bb}^h(\theta)$  given in (\ref{eq:final-estimator}) satisfies $\hat{\bb}^h = (\hat{\bb}_{I_0}^{\mathcal{O}}, \pmb 0_{I_0^c}) $ and $\|\hat{\bb}^h - \bb^*\|_2 \le  C_5\sqrt{\frac{\log p}{kn}}$.
 
\end{theorem}

Theorem \ref{Tm:3} guarantees that the estimator $\hat{\bb}^h$ in our proposed two-stage framework possesses estimation consistency and bi-level variable selection consistency, when conditions of Theorem \ref{Tm:2} hold and the condition of minimum individual signal strength is satisfied. Note that such signal strength condition is fairly mild and the bound can decrease arbitrarily closed to 0 with the growth of sample size $n$.
 
\section{Implementation} \label{sec:implementation}
We discuss the implementation of the proposed two-stage M-estimator framework in this section, including finding a stationary point in program  (\ref{eq:ex-group-penalization}) for a fixed $\lambda$ and the tuning parameters selection for $\lambda$ and $\theta$.

Note that the optimization in (\ref{eq:ex-group-penalization}) may not be a convex optimization problem since we allow both loss function $\mL_{n}$ and $\rho$ to be non-convex. To obtain the corresponding stationary point,  we use composite gradient descend algorithm \citep{nesterov2013gradient}. Recall $q_\lambda(\bb) =  \sum_{j=1}^J \sqrt{d_j}\lambda \|\bb_j\|_2 - \sum_{j=1}^{J} \rho(\|\bb_j\|_2,\sqrt{d_j}\lambda )$ and let $\bar{L}_{n}(\bb)=\mL_{n}(\bb)-q_\lambda(\bb)$. We can rewrite the program as
\begin{equation*}
\hat{\bb} \in \argmin_{\|\bb\|_1 \le R} \left\{ \bar{L}_n(\bb) + \sum_{j=1}^J \sqrt{d_j}\lambda \|\bb_j\|_2 \right\}.
\end{equation*}
Then the composition gradient iteration is given by
\begin{equation} \label{eq:interate}
\bb^{t+1} \in \argmin_{\|\bb\|_1 \le R} \left\{ \frac{1}{2} \| \bb - (\bb^t -  \frac{\nabla \bar{L}_n(\bb^t)}{\eta})\|_2^2 + \sum_{j=1}^J \lambda \eta  \sqrt{d_j}\|\bb_j\|_2   \right\},
\end{equation}
where $\eta >0$ is the step size for the update and can be determined by the backtracking line search method described in \cite{nesterov2013gradient}. A simple calculation shows that the iteration in (\ref{eq:interate}) takes the form
\begin{equation*}
\bb^{t+1}_j = S_{ \lambda \eta \sqrt{d_j}} \left(\left(\bb^t - \eta \nabla \bar{L}_n(\bb^t)\right)_j\right)
\end{equation*}
for $j=1, \cdots, J$, where $S_{\sqrt{d_j} \lambda  \eta}(\cdot)$ is the group soft-thresholding operator defined as 
\begin{equation*}
S_{\delta}(\bz) :=   \left(1 - \frac{\delta}{ \|\bz\|_2}\right)_+ \bz.
\end{equation*}
We further adopt the following two-step procedure discussed in \citet{loh2017statistical} to guarantee the convergence to a stationary point of the non-convex optimization problem in (\ref{eq:ex-group-penalization}).
\begin{enumerate}
\item[]{\it Step 1:} Run the composite gradient descent using a Huber loss function with convex group Lasso penalty to get an initial estimator.
\item[]{\it Step 2:}  Run the composite gradient descent on program (\ref{eq:ex-group-penalization}) using the initial estimator from Step 1.
\end{enumerate}

As to the tuning parameters selection, the optimal values of tuning parameters $\lambda$ and $\theta$ are chosen from a two-dimensional grid search using the cross-validation. In particular, the searching grid is formed by partitioning a rectangle uniformly in the scale of ($\theta,  \log(\lambda)$). Motivated by conditions of Theorem \ref{Tm:1} and Theorem \ref{Tm:3}, the range of the rectangle can be chosen as $C_{11}\sqrt{\frac{\log p}{n}} \le \lambda \le C_{12}\sqrt{\frac{\log p}{n}}$ and $C_{21}\sqrt{\frac{\log p}{kn}} < \theta \le C_{22}$. The optimal values are then found by the combination that minimizes the cross-validated trimmed mean squared prediction error. 

\begin{remark}
When the data are contaminated or heavy-tailed, the conventional mean squared prediction error for the cross-validation is not resistant to outliers in validation sets and may provide a biased selection of tuning parameters. In our simulation studies, we found that a robust measurement of prediction errors for the cross-validation, such as the trimmed mean squared error, or the error computed using the Huber loss or Cauchy loss, is robust to outliers in the validation set and yields better results in estimation and group/variable selection. For all numerical studies in this paper,  we adopt the trimmed mean squared prediction error for the sake of simplicity. As a comparison, we also report some simulation results using the mean squared prediction error for the cross-validation in Appendix.
\end{remark}

\section{Simulation Studies} \label{sec:simulation}
In this section, we assess the performance of our two-stage M-estimator framework by considering different types of loss functions and penalty functions through various models. The data are generated from the following model
\begin{equation*} 
   	 y_i=\bx_i^T\bb^*+\epsilon_i, \quad 1 \le i \le n.
\end{equation*}
The covariates vector $\bx_i$s are generated from a multivariate normal distribution with mean $\bf 0$ and covariance  $\bSigma$ independently. For covariance $\bSigma=(\sigma_{ij})_{p\times p}$, we choose

\begin{equation*}
\sigma_{ij}=\begin{cases}
1 & \text{if } i=j,\\
 (-1)^{i+j}a& \text{if } i\neq j \text{ and } i,j \text{ are in the same group}, \\
(-1)^{i+j} a b &   \text{if } i\neq j \text{ and } i,j  \text{ are in different groups},
\end{cases}
\end{equation*}
where $a=0.8$ or $0.5$ and $b=0.8$ or $0.5$. Let  $\bb^*=\bphi \cdot |\bb^*|$, where $\bphi$ is a $p$-dimensional vector with the $j$th element being $(-1)^{j+1}$.

\begin{example} \label{ex:1} {\bf (Group-level Sparsity)} The number of observations $n=100$ and  the number of variables $p=500$ with $J=100$ unequal-size groups. We choose $a=0.8$ and $b=0.5$. The model includes only between-group sparsity with five relevant groups,
$|\bb_1^*|=|\bb_2^*|=(\underbrace{3,\cdots,3}_4)^T=\bf {3}_4^{T}, \quad |\bb_3^*|=|\bb_4^*|=\bf {2}_6,\quad
 |\bb_5^*|=\bf {1.5}_5, \quad \bb_6^*=\cdots=\bb_{100}^*=\bf {0}_5.$
We generate random error $\epsilon_i$ from the following 3 scenarios: (a) $N(0,1)$, (b) $t_1$, (c) Mix Cauchy ($70\%$ are from $N(0,1)$ and $30\%$ are from standard Cauchy).
\end{example}

We consider bi-level penalized M-estimators with different types of loss functions (the $\ell_2$ loss, Huber loss, Cauchy loss) and two types of penalty functions (the Lasso and MCP penalties). In particular, we evaluate the performance of non-group estimators, one-stage estimators and two-stage estimators. Without causing any confusion, let $\hat{\bb}$ be any estimator of $\bb^*$. Its performances on both parameter estimation and group/variable selection   were evaluated by the following eight
measurements:
\begin{enumerate}
\item[(1)] $\ell_2$ error, which is defined as $\|\hat{\bb} - \bb^*\|_2$.
\item[(2)] $\ell_1$ error, which is defined as $\|\hat{\bb} - \bb^*\|_1$.
\item[(3)] Model size (MS), the average number of selected covariates.
\item[(4)] Group size (GS), the average number of selected groups.
\item[(5)] False positives rate for individual variable selection (FPR), the percent of selected covariates which are actually unimportant variables.
\item[(6)] False negatives rate  for individual variable selection (FNR), the percent of non-selected covariates which are actually important variables.
\item[(7)] False positives rate for group variable selection (GFPR), the percent of selected groups which are actually unimportant groups.
\item[(8)] False negatives rate for group variable selection (GFNR), the percent of non-selected groups which are actually important groups.
\end{enumerate}
Note that $\text{FPR}=\frac{|\hat{I} \bigcap I_0^c|}{|I_0^c|}\times 100\%$, $ \text{FNR}=\frac{|\hat{I}^c \bigcap I_0|}{|I_0|}\times 100\%$, $\text{GFPR}=\frac{|\hat{S} \bigcap S^c|}{|S^c|}\times 100\%$ and $\text{GFNR}=\frac{|\hat{S}^c \bigcap S|}{|S|}\times 100\%$, where $\hat{I}=\{m: \hat{\beta}_m \ne 0,~ 1 \le m \le p\}$, $I_0=\{m: \beta^*_m \ne 0, ~1 \le m \le p\}$, $\hat{S}=\{j: \hat{\bb}_j \ne \b0, ~ 1 \le j \le J\}$ and $S=\{j: \bb^*_j \ne \b0,~ 1 \le j \le J\}$.

 The model considered in Example \ref{ex:1} contains only the between-group sparsity. We also assess the performance of the two-stage M-estimator framework under models with bi-level sparsity in the following example.

\begin{example} \label{ex:2} {\bf (Bi-level Sparsity)} The number of observations $n=100$ and we generate the random error $\epsilon$ following the same three scenarios described in Example \ref{ex:1}.

\begin{itemize}
    \item[(i)] The number of variables $p=500$ with $J=100$ unequal-size groups. We choose $a=0.8$ and $b=0.5$. The model includes within-group sparsity among six relevant groups,
     $|\bb_1^*|=(1.5,2, 0, 2.5)^T$,  $|\bb_2^*|=(3,2,0,0,2)^T$, $|\bb_3^*|=(1.5,0,2.5, 3, 0,0)^T$, 
    $|\bb_4^*|=( 2,1.5,\underbrace{0,\cdots,0}_4)^T$, $|\bb_5^*|=(2.5,0,0,0)^T$, 
    $|\bb_6^*|=(3, 2.5, 2.5, 2, 1.5)^T$, $\bb_7^*=\cdots=\bb_{100}^*=(\underbrace{0,\cdots,0}_{5})^T$.
    \item[(ii)] Similar to (i) except that we choose $a=0.5$ and $b=0.8$.
    
    \item [(iii)] The number of variables $p=1000$ with $J=100$ unequal-size groups. We choose $a=0.8$ and $b=0.5$. The model includes within-group sparsity in among four relevant groups, $|\bb_1^*|=(3,2, 0, 0, 0)^T$, \\
    $|\bb_2^*|=(1.5, 2, 2.5, 2.5, 3, \underbrace{0,\cdots,0}_5)^T$,
    $|\bb_3^*|=(1.5, 0, 2.5, 3, 0, 3, 2, 1.5, \underbrace{0,\cdots,0}_7)^T$,
    $|\bb_4^*|=(3,3,2.5,2.5,2,2,1.5, 1.5, 1.5, 1.5)^T$, 
    $\bb_5^*=\cdots=\bb_{100}^*=(\underbrace{0,\cdots,0}_{10})^T$.
\end{itemize}
\end{example}
Finally, we design a simulation setting to evaluate the performance of the two-stage M-estimator framework when covariates are contaminated or not sub-Gaussian.

\begin{example} \label{ex:3} {\bf (Contamination on $\bx$)} All the settings are similar to Example \ref{ex:2}(i), except that we let $n=120$ and covariates be partially contaminated after the data generation. In particular, $20\%$ of the observations in $\bx$ are replaced by data generated from $\chi^2(10)$ first, and then recentered to have mean zero.

\end{example}

We ran 100 simulations for each scenario described in Examples \ref{ex:1}-\ref{ex:3}. While fixing $v(\bx)\equiv w(\bx) \equiv 1$ for Examples \ref{ex:1} and \ref{ex:2}, we consider the general two-stage M-estimator framework with $v(\bx)\equiv 1$ and $w(\bx)=\min\left\{1,\frac{4} {\|\bx\|_{\infty}}\right\}$ in Example \ref{ex:3}. As introduced in Section \ref{sec:implementation}, we choose two tuning parameters $\lambda$ and $\theta$ optimally with $10$-fold cross-validation, with $\lambda$ ranging in ($0.01\sqrt{\frac{\log p}{n}}, 10\sqrt{\frac{\log p}{n}}$) and $\theta$ ranging in ($0.01\sqrt{\frac{\log p}{kn}}, 0.5$). The results from  Example \ref{ex:1} to \ref{ex:3} are reported in Table \ref{tb:ex1} to \ref{tb:ex3}, respectively. 
Note that we consider the one-stage estimators with the Lasso penalty as the GLasso-type estimators. For the MCP penalty, we call the corresponding non-group estimators, one-stage estimators and two-stage estimators the MCP-type, GMCP-type and GMCP-HT-type estimators, respectively.


\begin{table}[thp]

\begin{tabular}{|cl|rrrrrrr|}
\hline
\multicolumn{2}{|l|}{\multirow{2}{*}{}}      & \multicolumn{3}{c}{Group Lasso}        &                      & \multicolumn{3}{c|}{Group MCP}          \\ \cline{3-5} \cline{7-9} 
\multicolumn{2}{|l|}{}      & LS         & Huber       & Cauchy      &                      & LS           & Huber       & Cauchy     \\ \hline
\multirow{8}{*}{N(0,1)}     & $\ell_2$ error & 1.27 & 1.29 & 1.39  &                      & 0.92   & 0.93  & 0.95 \\
                            & $\ell_1$ error & 6.3        & 6.59        & 6.93        &                      & 3.75         & 3.77        & 3.85       \\
                            & MS             & 55.9       & 66.21       & 66.01       &                      & 31.43        & 33.23       & 33.09      \\
                            & GS             & 11.18      & 13.24       & 13.2        &                      & 6.29         & 6.65        & 6.62       \\
                            & FPR            & 6.51       & 8.69        & 8.73        &                      & 1.35         & 1.73        & 1.7        \\
                            & FNR            & 0          & 0.36        & 1.76        & \multicolumn{1}{l}{} & 0            & 0           & 0          \\
                            & GFPR           & 6.51       & 8.69        & 8.73        &                      & 1.36         & 1.74        & 1.71       \\
                            & GFNR           & 0          & 0.4         & 1.8         & \multicolumn{1}{l}{} & 0            & 0           & 0          \\ \hline
\multirow{8}{*}{$t_1$}      & $\ell_2$ error & 13.77      & 2.02        & 1.82        &                      & 24.96        & 2.72        & 2.46       \\
                            & $\ell_1$ error & 166.82     & 11.04       & 9.59        &                      & 243.53       & 10.88       & 10.22      \\
                            & MS             & 114.89     & 71.75       & 70.12       &                      & 65.64        & 27.9        & 29.15      \\
                            & GS             & 23         & 14.35       & 14.03       &                      & 13.16        & 5.58        & 5.83       \\
                            & FPR            & 19.26      & 9.94        & 9.59        &                      & 9.11         & 0.61        & 0.87       \\
                            & FNR            & 6.32       & 1.8         & 1.68        & \multicolumn{1}{l}{} & 10.6         & 0           & 0          \\
                            & GFPR           & 19.26      & 9.94        & 9.59        &                      & 9.12         & 0.61        & 0.87       \\
                            & GFNR           & 6          & 1.8         & 1.6         & \multicolumn{1}{l}{} & 10           & 0           & 0          \\ \hline
\multirow{8}{*}{Mix Cauchy} & $\ell_2$ error & 12.84      & 1.42        & 1.36        &                      & 16.92        & 1.46        & 1.36       \\
                            & $\ell_1$ error & 178.11     & 7.44        & 6.92        &                      & 225.05       & 5.82        & 5.48       \\
                            & MS             & 94.6       & 72.11       & 71.25       &                      & 51.99        & 27.2        & 29.45      \\
                            & GS             & 18.92      & 14.42       & 14.26       &                      & 10.4         & 5.44        & 5.89       \\
                            & FPR            & 14.75      & 9.94        & 9.79        &                      & 5.84         & 0.46        & 0.94       \\
                            & FNR            & 1.8        & 0.36        & 1           & \multicolumn{1}{l}{} & 3.04         & 0           & 0          \\
                            & GFPR           & 14.75      & 9.94        & 9.8         & \multicolumn{1}{l}{} & 5.84         & 0.46        & 0.94       \\
                            & GFNR           & 1.8        & 0.4         & 1           &                      & 3            & 0           & 0          \\ \hline
\end{tabular}
\caption{Simulation results under the model with only between-group sparsity in Example \ref{ex:1}. The mean $\ell_2$ error, $\ell_1$ error, MS, GS, FPR (\%), FNR(\%), GFPR (\%) and GFNR (\%) out of 100 iterations are displayed. }
\label{tb:ex1}
\end{table}

\begin{table}[thp]

\begin{tabular}{|cl|llllllll|}
\hline
\multicolumn{2}{|l|}{\multirow{2}{*}{}} & \multicolumn{2}{c}{MCP} &  & \multicolumn{2}{c}{GMCP} &  & \multicolumn{2}{c|}{GMCP-HT} \\ \cline{3-4} \cline{6-7} \cline{9-10} 
\multicolumn{2}{|l|}{} & Huber & Cauchy &  & Huber & Cauchy &  & Huber & Cauchy \\ \hline
\multirow{8}{*}{N(0,1)} & $\ell_2$ error & 6.98 & 7.45 &  & 1.67 & 1.66 &  & 1.57 & 1.6 \\
 & $\ell_1$ error & 29.3 & 31.52 &  & 7.5 & 7.49 &  & 6.79 & 6.9 \\
 & MS & 28.77 & 27.56 &  & 53.89 & 53.55 &  & 30.49 & 30.54 \\
 & GS & 18.75 & 18 &  & 10.79 & 10.71 &  & 8.27 & 8.21 \\
 & FPR & 3.54 & 3.37 &  & 7.64 & 7.57 &  & 2.79 & 2.8 \\
 & FNR & 31.41 & 33.53 &  & 0 & 0 &  & 0 & 0 \\
 & GFPR & 13.66 & 12.8 &  & 5.1 & 5.01 &  & 2.41 & 2.35 \\
 & GFNR & 1.5 & 0.5 &  & 0 & 0 &  & 0 & 0 \\ \hline
\multirow{8}{*}{$t_1$} & $\ell_2$ error & 11.31 & 11.5 &  & 4.37 & 3.75 &  & 4.34 & 3.68 \\
 & $\ell_1$ error & 46.5 & 47.37 &  & 19.45 & 16.68 &  & 19.28 & 16.08 \\
 & MS & 12.61 & 9.85 &  & 52.57 & 47.53 &  & 34.7 & 31.44 \\
 & GS & 9.99 & 8.51 &  & 10.51 & 9.5 &  & 8.28 & 7.74 \\
 & FPR & 1.28 & 0.83 &  & 7.37 & 6.33 &  & 3.72 & 3.03 \\
 & FNR & 62.29 & 65.53 &  & 0.18 & 0.18 &  & 1.59 & 1.12 \\
 & GFPR & 4.94 & 3.4 &  & 4.83 & 3.76 &  & 2.51 & 1.9 \\
 & GFNR & 10.83 & 11.5 &  & 0.5 & 0.5 &  & 1.33 & 0.83 \\ \hline
\multirow{8}{*}{Mix Cauchy} & $\ell_2$ error & 8.9 & 8.91 &  & 2.47 & 2.11 &  & 2.39 & 2.03 \\
 & $\ell_1$ error & 37.94 & 38.24 &  & 11.14 & 9.42 &  & 10.29 & 8.62 \\
 & MS & 18.87 & 18.91 &  & 47.4 & 48.94 &  & 29.24 & 29.06 \\
 & GS & 13.84 & 13.97 &  & 9.48 & 9.79 &  & 7.24 & 7.42 \\
 & FPR & 2.03 & 2.08 &  & 6.29 & 6.61 &  & 2.54 & 2.5 \\
 & FNR & 46.65 & 48 &  & 0 & 0 &  & 0.06 & 0 \\
 & GFPR & 8.48 & 8.59 &  & 3.7 & 4.03 &  & 1.32 & 1.51 \\
 & GFNR & 2.17 & 1.67 &  & 0 & 0 &  & 0 & 0 \\ \hline
\end{tabular}
\caption{Simulation results under the model with bi-level sparsity in Example \ref{ex:2}(i). The mean $\ell_2$ error, $\ell_1$ error, MS, GS, FPR (\%), FNR(\%), GFPR (\%) and GFNR (\%) out of 100 iterations are displayed.}
\label{tb:ex2.1}
\end{table}

\begin{table}[thp]

\begin{tabular}{|cl|rrrrrrr|}
\hline
\multicolumn{2}{|l|}{\multirow{2}{*}{}}      & \multicolumn{3}{c}{GMCP - HT} &                      & \multicolumn{3}{c|}{WGMCP - HT} \\ \cline{3-5} \cline{7-9} 
\multicolumn{2}{|l|}{}      & LS        & Huber   & Cauchy  &                      & LS         & Huber   & Cauchy   \\ \hline
\multirow{8}{*}{N(0,1)}     & $\ell_2$ error & 7.49      & 7.52    & 7.54    &                      & 6.74       & 6       & 4.97     \\
                            & $\ell_1$ error & 43.56     & 42.78   & 40.76   &                      & 35.75      & 28.81   & 22.42    \\
                            & MS             & 71.76     & 66.93   & 54.81   &                      & 60.82      & 38.22   & 32.53    \\
                            & GS             & 17.34     & 16.26   & 13.74   &                      & 14.02      & 9.38    & 9.72     \\
                            & FPR            & 11.64     & 10.63   & 8.15    &                      & 9.28       & 4.6     & 3.38     \\
                            & FNR            & 8.59      & 8.35    & 9.06    & \multicolumn{1}{l}{} & 5.82       & 5.88    & 4.76     \\
                            & GFPR           & 12.72     & 11.59   & 8.9     &                      & 9.04       & 3.98    & 4.33     \\
                            & GFNR           & 10.33     & 10.5    & 10.5    & \multicolumn{1}{l}{} & 8          & 6       & 5.83     \\ \hline
\multirow{8}{*}{$t_1$}      & $\ell_2$ error & 125.26    & 8.46    & 8.54    &                      & 126.92     & 6.96    & 6.43     \\
                            & $\ell_1$ error & 2081.98   & 47.63   & 46.78   &                      & 2099.27    & 33.87   & 30.84    \\
                            & MS             & 96.16     & 61.26   & 52.46   &                      & 86.38      & 37.42   & 35       \\
                            & GS             & 22.76     & 15.2    & 14.12   &                      & 19.26      & 9.07    & 9.56     \\
                            & FPR            & 17.39     & 9.71    & 7.91    &                      & 15.35      & 4.57    & 4.05     \\
                            & FNR            & 28.29     & 15.65   & 16.12   & \multicolumn{1}{l}{} & 27.94      & 9.71    & 9.18     \\
                            & GFPR           & 19.76     & 10.93   & 9.77    &                      & 16.06      & 4.02    & 4.56     \\
                            & GFNR           & 30.17     & 17.83   & 17.67   & \multicolumn{1}{l}{} & 30.67      & 11.83   & 12.17    \\ \hline
\multirow{8}{*}{Mix Cauchy} & $\ell_2$ error & 18.52     & 7.72    & 7.66    &                      & 18.48      & 5.97    & 5.22     \\
                            & $\ell_1$ error & 211.8     & 43.23   & 39.92   &                      & 210.62     & 28.68   & 24.67    \\
                            & MS             & 75.62     & 64.31   & 48.28   &                      & 63.31      & 37.8    & 35.46    \\
                            & GS             & 18.17     & 15.69   & 12.39   &                      & 13.98      & 9.41    & 10.16    \\
                            & FPR            & 12.69     & 10.19   & 6.87    &                      & 10.02      & 4.52    & 4        \\
                            & FNR            & 15.76     & 11.18   & 11.29   & \multicolumn{1}{l}{} & 12.29      & 6       & 4.94     \\
                            & GFPR           & 14.09     & 11.16   & 7.59    & \multicolumn{1}{l}{} & 9.41       & 4.09    & 4.74     \\
                            & GFNR           & 17.83     & 13.33   & 12.33   &                      & 14.5       & 7.17    & 5        \\ \hline
\end{tabular}
\caption{Simulation results under the model with 20\% contamination on X in Example \ref{ex:3}. The mean $\ell_2$ error, $\ell_1$ error, MS, GS, FPR (\%), FNR(\%), GFPR (\%) and GFNR (\%) out of 100 iterations are displayed. }
\label{tb:ex3}
\end{table}

We mainly evaluate the performance of one-stage estimators in Example \ref{ex:1} since there only exists the between-group sparsity. Table \ref{tb:ex1} shows that with the same loss function, while the GMCP-type estimators perform comparably to the GLasso-type estimators in the estimation, the former has better group/variable selection accuracy than the latter. This is consistent with the group oracle property stated in Theorem \ref{Tm:2}. As expected, for the estimators with the same penalty, while they behave similarly in the light-tailed setting ($N(0,1)$), estimators using Huber loss and Cauchy loss largely outperform the least squares estimator for the heavy-tailed settings ($t_1$ and Mix Cauchy). 


We compare the results of non-group estimators, one-stage estimators and two-stage estimators for Example \ref{ex:2}. Note that here we only consider the MCP penalty since it has been shown to perform better than the Lasso penalty. We also omit the results of the least squares estimators since they are not robust to the heavy-tailed settings. For Example \ref{ex:2}(i), Table \ref{tb:ex2.1} shows that the GMCP-type estimators outperform the MCP-type estimators in all measurements, as the former incorporates the grouping structure in $\bx$. By comparing the results of GMCP-type estimators and GMCP-HT-type estimators, we see that the extra hard-thresholding step in the two-stage estimators can effectively improve the estimation and group/variable selection performance. 
In addition, estimators using the Cauchy loss further outperform the one with Huber Loss for the heavy-tailed settings,  showing that the re-descending estimators are more robust to outliers and more efficient for irregular settings. We observe similar patterns in the results of Examples \ref{ex:2}(ii)-(iii) and thus we omit those results in this paper.

In Example \ref{ex:3} we only compare the performance of two-stage estimators with their weighted version. Table \ref{tb:ex3} indicates that the two-stage estimators with well-chosen $w(\bx)$ perform better in all cases than the two-stage estimators with $w(\bx)=1$. Again when the errors are heavy-tailed ($t_1$ and Mix Cauchy), the least squares estimator loses its efficiency and the re-descending estimators produced by Cauchy loss perform the best for all scenarios. 

In summary, our simulation studies show that in the proposed two-stage M-estimator framework, (1) the GP Stage can utilize the grouping structure to yield satisfactory parameter estimation and group variable selection results for irregular settings, if a robust loss function (e.g. Huber loss and Cauchy loss) is used; (2) the HT Stage further improve the performance by filtering out the non-important selected variable from the first stage; (3) the two-stage M-estimators with re-descending loss functions (e.g. Cauchy loss) and concave folded penalties consistently render more satisfactory results when data are heavy-tailed or strongly contaminated ($t_1$ and Mix Cauchy).

\section{Real Data Example}
In this section, we use the NCI-60 data, a gene expression dataset collected from Affymetrix HG-U133A chip, to illustrate the performance of the proposed two-stage penalized M-estimators evaluated in Section {\ref{sec:simulation}}. The NCI-60 data consist of data on 60 human cancer cell lines and can be downloaded via the web application CellMiner (\url{http://discover.nci.nih.gov/cellminer/}). The study is to predict the protein expression on the KRT18 antibody from other gene expression levels. The expression level of the protein {\it keratin} 18 is known to be persistently expressed in carcinomas (\cite{oshima1996oncogenic}). After removing the missing data, there are $n=59$ samples with  $21,944$ genes in the dataset. One can refer \cite{shankavaram2007transcript} for more details.

We first perform some pre-screenings by keeping only 2000
genes with the largest variations and choosing 500 genes out of which are most correlated with the response variable. Then for each gene, we use B-spline with 5 bases to form a group with 5 variables. Thus our final dataset has $n=59$ samples, $p=2500$ variables and $J=500$ groups. Similar to our simulation studies, we apply the non-group estimators, one-stage estimators and two-stage estimators to select important genes, with tuning parameter $\lambda$ and $\theta$ chosen from the $10$-folded cross-validation with $\lambda$ ranging in ($0.01\sqrt{\frac{\log p}{n}}, 10\sqrt{\frac{\log p}{n}}$) and $\theta$ ranging in ($0.01, 1$). We also control the numbers of selected variables to be less than 75\% of the sample size, in order to avoid overfitting.   We report results from six methods: Huber-MCP, Cauchy-MCP, Huber-GMCP, Cauchy-GMCP, Huber-GMCP-HT, Cauchy-GMCP-HT.

\begin{figure}[thp]
  	\centering{
  	\makebox[\textwidth]{\includegraphics[width=\textwidth]{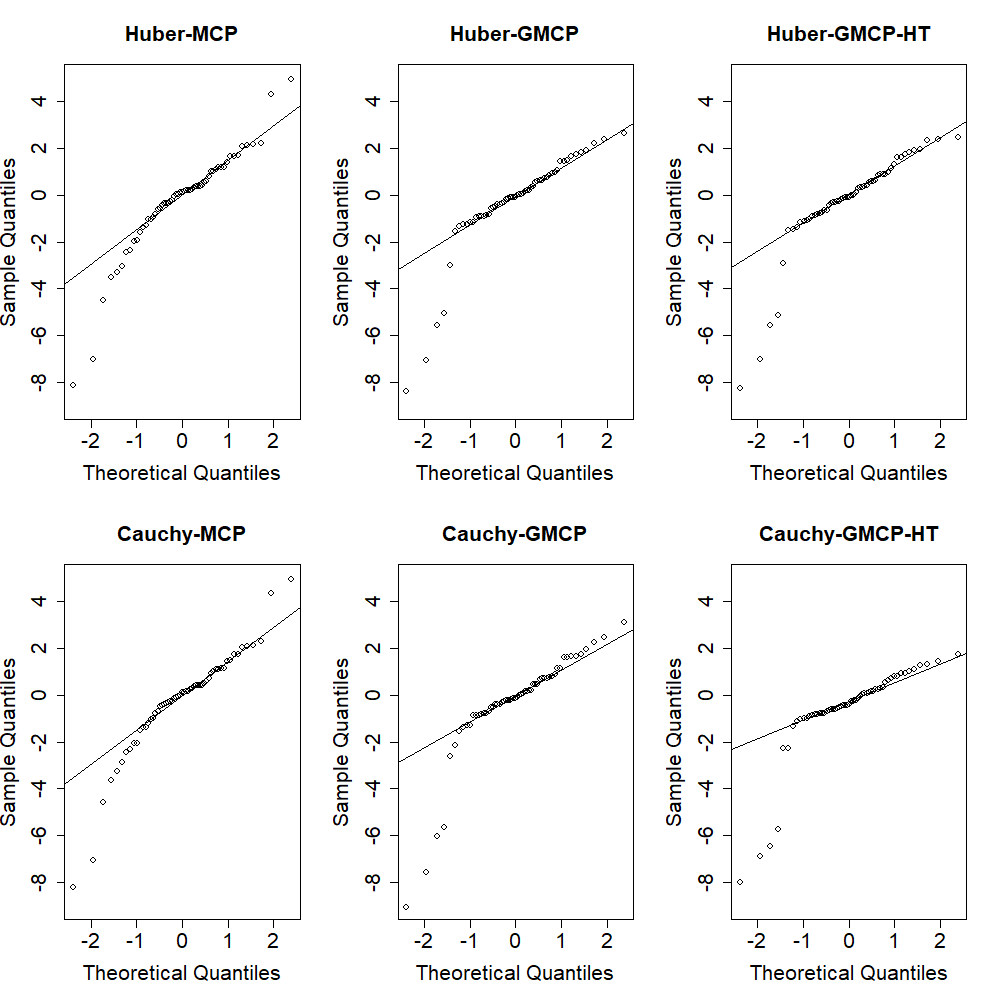}}
  	 \caption{QQ plots of the residuals from Huber-MCP, Cauchy-MCP, Huber-GMCP, Cauchy-GMCP, Huber-GMCP-HT, Cauchy-GMCP-HT.}
  	 \label{fig:resid}
  	 }
 \end{figure}

The QQ-plots of the residuals generated from these six methods are shown in Figure \ref{fig:resid}. It can be seen that each residual distribution has a longer tail on the left side, implying that the data may be contaminated or heavy-tailed. Table \ref{tb:vs} displays the selected important genes. The number of selected genes from these methods are 9 (Huber-MCP), 3 (Huber-GMCP), 3 (Huber-GMCP-HT), 9 (Cauchy-MCP), 2 (Cauchy-GMCP) and 5 (Cauchy-GMCP-HT), respectively. It is not surprising that gene KRT8 is selected by all six methods due to its largest correlation with the response variable and the long history of being paired with KRT18 in cancer studies \citep{li2016silac, Walker2007cancer}. Notice that gene INHBB is singled out by the four methods that incorporate grouping information. This gene has been shown to play essential roles in tumorigenesis and migration \citep{kita2017activin, wijayarathna2016activins}. Table \ref{tb:vs} also shows that the Huber-MCP and Cauchy-MCP both select the same genes, which indicates that the contamination in the data may not be strong enough to cause different selection results between these two loss functions. In addition, it is reasonable to observe that the Huber-GMCP and Huber-GMCP-HT also select exactly the same genes, since there is no sparsity within each group in the data.  It is also worth mentioning that the Cauchy-GMCP-HT selects more genes than the Cauchy-GMCP. It means that the chosen $\lambda$ of Cauchy-GMCP-HT is different from that of Cauchy-GMCP, as for the former the optimal $\lambda$ and $\theta$ are found together by the cross-validation.


  \begin{table}[thp]
 \caption{\label{tb:vs}Selected genes by Huber-MCP, Cauchy-MCP, Huber-GMCP, Cauchy-GMCP, Huber-GMCP-HT, Cauchy-GMCP-HT.}
 \centering
\begin{tabular}{|c|ccccc|}
\hline
\multirow{2}{*}{Huber-MCP}  & KRT8 & NRN1   & KRT18  & GAS7   & EPS8L2 \\
                            & GPX3 & TRIM29 & LAD1   & SEMA5A &        \\ \hline
Huber-GMCP                  & KRT8 & INHBB  & PBX1   &        &        \\ \hline
Huber-GMCP-HT               & KRT8 & INHBB  & PBX1   &        &        \\ \hline
\multirow{2}{*}{Cauchy-MCP} & KRT8 & NRN1   & KRT18  & GAS7   & EPS8L2 \\
                            & GPX3 & TRIM29 & LAD1   & SEMA5A &        \\ \hline
Cauchy-GMCP                 & KRT8 & INHBB  &        &        &        \\ \hline
Cauchy-GMCP-HT              & KRT8 & NR2F2  & NOTCH3 & INHBB  & SIRPA  \\ \hline
\end{tabular}
\end{table} 

\begin{figure}[thp]
  \makebox[\textwidth]{\includegraphics[width=\textwidth]{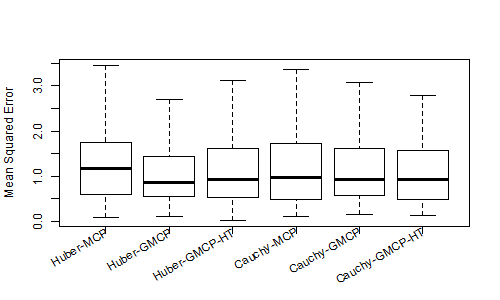}}
  \caption{Boxplot of the Mean Squared Error of predictions.}
  \label{fig:boxplot}
 \end{figure}

For further investigation, we randomly choose 6 observations as the test set and applied those six methods to the rest patients to get the estimation of the coefficients, then compute the prediction error on the test set. We repeat the random splitting 100 times and the boxplots of the Mean Squared Error of predictions are shown in Figure \ref{fig:boxplot}. It can be seen that the Huber-GMCP and Cauchy-GMCP perform better than the other methods. This is not surprising since there is only between-group sparsity in the dataset. In addition, Figure \ref{fig:boxplot} also shows that Cauchy-type estimators perform similarly to the corresponding Huber-type estimators, which indicates that when there exists only moderate contamination in the data, it may be sufficient to consider the convex Huber loss in our framework.

\section{Discussion}
Bi-level variable selection and parameter estimation are crucial when covariates function group-wisely in high dimensional settings. It has become even more challenging when data are contaminated or heavy-tailed. In this paper, we propose a two-stage penalized M-estimator framework for high-dimensional bi-level variable selection. This framework consists of two stages:  penalized M-estimation with a concave $\ell_2$-norm penalty achieving the consistent group selection at the first stage, and a post-hard-thresholding operator to achieve the within-group sparsity at the second stage. The proposed framework is very general in that it covers a wide range of loss functions and penalty functions, allowing both functions to be non-convex. Thus if the data are strongly contaminated, either in covariates or random error, we are still able to perform bi-level variable selection efficiently through the proposed framework.

Theoretically, we establish statistical properties of the proposed two-stage penalized M-estimator in ultra high-dimensional settings when $p$ grows with $n$ at an almost exponential rate. In particular, for the estimator at the Group Penalization Stage, we show its local estimation consistency at the minimax rate enjoyed by LS-GLasso and establish the local group selection consistency. For the post-hard-thresholding estimator at the second stage, we show that it naturally inherits all those nice statistical properties from the first stage and further possesses bi-level variable selection consistency.
These theoretical results require weak assumptions on model settings and are applicable even though the random error and  covariates are  heavy-tailed  or the dataset is contaminated by outliers.

Our framework is computationally efficient, and is able to find a well-behaved local stationary point if a consistent initial such as Huber group Lasso is used. Our numerical studies show satisfactory finite sample performances of the two-stage penalized M-estimator under different irregular settings, which is consistent with our theoretical findings. In particular at the first stage, among some of the possible choices of loss and penalty functions that fit in the proposed framework, our numerical studies suggest considering a re-descending loss function, such as Cauchy loss or Tukey's biweight loss, with a group concave folded penalty, such as group MCP penalty, when the data are strongly  contaminated.

\appendix
\section{Appendix}
\subsection{Mean squared prediction error for the cross-validation}

Compared with the mean squared prediction error for the cross-validation,  the trimmed version is robust to outliers in validation sets and provides a better selection of tuning parameters. To illustrate this point, we re-run the simulation using the mean squared prediction error for the cross-validation. We report the results for Example \ref{ex:2}(i) in Table \ref{rtb:ex2.1_ms}. Note that the results with the trimmed mean squared prediction error are displayed in Table \ref{tb:ex2.1} in our paper.
In the light-tailed setting ($N(0,1)$), with similar estimation errors, it is not surprising that the mean squared prediction error for the cross-validation yields slightly better group/variable selection performance than the trimmed version, as there are not any outliers in the dataset.  However, in the heavy-tailed settings ($t_1$ and Mix Cauchy),  we can clearly see that the robust GMCP and GMCP-HT using the trimmed mean squared prediction error perform better in both parameter estimation and group/variable selection. In particular, the robust GMCP-HT method with the trimmed version is able to largely reduce the false negative rates in group/variable selection while maintaining competitive false positive rates.

\begin{table}[thp]
\begin{tabular}{|cl|llllllll|}
\hline
\multicolumn{2}{|l|}{\multirow{2}{*}{}} & \multicolumn{2}{c}{MCP} &  & \multicolumn{2}{c}{GMCP} &  & \multicolumn{2}{c|}{GMCP-HT} \\ \cline{3-4} \cline{6-7} \cline{9-10}
\multicolumn{2}{|l|}{} & Huber & Cauchy &  & Huber & Cauchy &  & Huber & Cauchy \\ \hline
\multirow{8}{*}{N(0,1)} & $\ell_2$ error & 7.23 & 7.34 &  & 1.69 & 1.67 &  & 1.58 & 1.57 \\
 & $\ell_1$ error & 30.35 & 30.85 &  & 7.55 & 7.49 &  & 6.81 & 6.74 \\
 & MS & 24.66 & 22.58 &  & 39.76 & 38.82 &  & 29.24 & 28.82 \\
 & GS & 16.79 & 15.44 &  & 8.96 & 8.73 &  & 7.7 & 7.61 \\
 & FPR & 2.77 & 2.41 &  & 4.71 & 4.52 &  & 2.53 & 2.45 \\
 & FNR & 33.71 & 35.76 &  & 0 & 0 &  & 0 & 0 \\
 & GFPR & 11.56 & 10.11 &  & 3.15 & 2.9 &  & 1.81 & 1.71 \\
 & GFNR & 1.33 & 1 &  & 0 & 0 &  & 0 & 0 \\ \hline
\multirow{8}{*}{$t_1$} & $\ell_2$ error & 11.33 & 11.36 &  & 5.15 & 4.53 &  & 4.99 & 4.44 \\
 & $\ell_1$ error & 46.55 & 47.09 &  & 22.72 & 19.32 &  & 22.32 & 19.47 \\
 & MS & 11.33 & 10.65 &  & 37.31 & 34.87 &  & 32.02 & 31.33 \\
 & GS & 9.16 & 9.11 &  & 7.73 & 7.14 &  & 8.17 & 8.73 \\
 & FPR & 1.05 & 0.92 &  & 4.38 & 3.82 &  & 3.35 & 3.17 \\
 & FNR & 63.24 & 63.59 &  & 4.88 & 3.53 &  & 6.71 & 5.71 \\
 & GFPR & 4.22 & 4.09 &  & 2.34 & 1.56 &  & 2.85 & 3.36 \\
 & GFNR & 13.5 & 12.17 &  & 7.83 & 5.5 &  & 8.5 & 7.17 \\ \hline
\multirow{8}{*}{MixCauchy} & $\ell_2$ error & 8.65 & 9.14 &  & 2.92 & 2.73 &  & 2.84 & 2.7 \\
 & $\ell_1$ error & 36.59 & 38.91 &  & 12.95 & 11.82 &  & 12.4 & 11.35 \\
 & MS & 19.17 & 16.15 &  & 35.32 & 34.46 &  & 27.69 & 26.63 \\
 & GS & 13.8 & 12.56 &  & 7.38 & 7.24 &  & 7.21 & 7.09 \\
 & FPR & 2.06 & 1.62 &  & 3.83 & 3.69 &  & 2.28 & 2.1 \\
 & FNR & 45.76 & 51 &  & 1 & 2.24 &  & 2 & 2.88 \\
 & GFPR & 8.44 & 7.12 &  & 1.56 & 1.51 &  & 1.45 & 1.4 \\
 & GFNR & 2.17 & 2.17 &  & 1.5 & 3 &  & 2.5 & 3.83 \\ \hline
\end{tabular}
\caption{Simulation results under the model with bi-level sparsity in Example \ref{ex:2}(i), with the mean squared prediction error for the cross-validation.  The mean $\ell_2$ error, $\ell_1$ error, MS, GS, FPR (\%), FNR(\%), GFPR (\%) and GFNR (\%) out of 100 iterations are displayed.}
\label{rtb:ex2.1_ms}
\end{table}

\subsection{Proofs}\label{proof} 

To prove Theorem \ref{Tm:1}, we need the following Lemma \ref{Lemma: gradient_bound}.
\begin{lemma} \label{Lemma: gradient_bound}
Suppose $\mL_{n}$ in (\ref{eq:gloss}) satisfies Assumption \ref{as:loss} and the random errors and covariates satisfy Assumption \ref{as:x}. For any $t \in (0,n)$, we have
\begin{equation*}
    \|\nabla\mL_{n}(\bb^*)\|_\infty \le C_0 \sqrt{\frac{t}{n}}
\end{equation*}
with probability at least $1-2p\exp(-t)$.
\end{lemma}
\textit{Proof.} The gradient of $\mL_{n}$ is
\begin{equation*}
\begin{split}
\nabla \mL_{n}(\bb^*)=&-\frac{1}{n} \sum_{i=1}^{n} w(\bx_i)\bx_il'(\epsilon_iv(\bx_i)).
\end{split}
\end{equation*}
If Assumption \ref{as:x}(ii) (a) holds, then
\begin{equation} \label{eq:em}
    \begin{split}
        E[w(\bx_i)\bx_il'(\epsilon_iv(\bx_i))]&=E[w(\bx_i)\bx_il'(\epsilon_i)]\\
        &=E[w(\bx_i)\bx_i] E[l'(\epsilon_i)]\\
        &=\b0,
    \end{split}
\end{equation}
where the second equality follows from  $\epsilon_i \indep \bx_i$. If Assumption \ref{as:x}(ii) (b) is satisfied instead, we obtain
\begin{equation}
    E[w(\bx_i)\bx_il'(\epsilon_iv(\bx_i))]=E[w(\bx_i)\bx_i E[l'(\epsilon_iv(\bx_i))|\bx_i] = \b0.
\end{equation}
Therefore,  $E[\nabla \mL_{n}(\bb^*)]=\b0$ under Assumption \ref{as:x}(ii). 

Let $\mu_j = E[w(\bx_i)x_{ij}]$, $j=1,2,\dots, p$. Then we have
\begin{equation} \label{eq:boundwxm}
    \begin{split}
        E|w(\bx_i)x_{ij}|^m=& E|w(\bx_i)x_{ij}-\mu_j+\mu_j|^m\\
        \le & E[2^{m-1}(|w(\bx_i)\bx_{ij}-\mu_j|^m+|\mu_j|^m)]\\
        \le & 2^{m-1} [E|w(\bx_i)x_{ij}-\mu_j|^m + \tau^m]\\
        \le & 2^{m-1} [m(\sqrt{2})^{m}k_0^m \Gamma(\frac{m}{2}) + \tau^m],
    \end{split}
\end{equation}
where $\max_{1 \le j \le p}|\mu_j| < \tau <\infty$ and the last inequality follows from Assumption \ref{as:x}(i), by which $w(\bx_i)x_{ij}$ is sub-Gaussian hence for $m > 0$ (\cite{rivasplata2012subgaussian})
\begin{equation*}\label{eq:gaussian}
E|w(\bx_i)x_{ij}-\mu_j|^m \le m(\sqrt{2})^{m}k_0^m\Gamma(\frac{m}{2}).
\end{equation*}

Next we bound the $E|w(\bx_i)x_{ij}l'(\epsilon_iv(\bx_i))|^m$ from the above. By Assumption \ref{as:loss}(i) and the bound in (\ref{eq:boundwxm}), we have
\begin{equation} \label{eq:boundlossm}
    \begin{split}
        E|w(\bx_i)x_{ij}l'(\epsilon_iv(\bx_i))|^m & \le k_1^m E|w(\bx_i)x_{ij}|^m \\
        & \le k_1^m 2^{m-1} [m(\sqrt{2})^{m}k_0^m \Gamma(\frac{m}{2}) + \tau^m].
    \end{split}
\end{equation}
By taking $m=2$ in (\ref{eq:boundlossm}), we obtain
\begin{equation} \label{eq:boundloss2}
    E|w(\bx_i)x_{ij}l'(\epsilon_iv(\bx_i))|^2 \le l_1, 
\end{equation}
where $l_1=k_1^2(8k_0^2+2\tau^2)$.
For all integer $m\ge3$, by equation (\ref{eq:boundlossm}) we have
\begin{equation}
    \begin{split}
        E|w(\bx_i)x_{ij}l'(\epsilon_iv(\bx_i))|^m & \le  k_1^m 2^{m-1} [m(\sqrt{2})^{m}k_0^m \Gamma(\frac{m}{2}) + \tau^m]\\
        & \le \frac{m!}{2} k_1^{m-2} (2\tau + \sqrt{2}k_0)^{m-2}[k_1^2(8k_0^2+2\tau^2)]\\
        & =\frac{m!}{2} l_2^{m-2}l_1,
    \end{split}
\end{equation}
where $l_2=k_1(2\tau + \sqrt{2}k_0)$. By Bernstein inequality (Proposition 2.9 of \citet{massart2007concentration}) we have
\begin{equation*}
P(|\frac{1}{n}\sum_{i=1}^{n} w(\bx_i)x_{ij}l'(\epsilon_iv(\bx_i))-\frac{1}{n}\sum_{i=1}^{n} E[w(\bx_i)x_{ij}l'(\epsilon_iv(\bx_i))]| \ge  \sqrt{\frac{2l_1t}{n}}+\frac{l_2t}{n})\le 2\exp(-t).
\end{equation*}
Together with equation (\ref{eq:em}), we have
\begin{equation*}
    P(|\frac{1}{n}\sum_{i=1}^{n} w(\bx_i)x_{ij}l'(\epsilon_iv(\bx_i))| \ge  C_0 \sqrt{\frac{t}{n}})\le 2\exp(-t)
\end{equation*}
for $t \in (0,n]$, where $C_0=\sqrt{2l_1}+l_2$. It then follows from union inequality that 
\begin{equation*}
    P(\|\nabla\mL_{n}(\bb^*)\|_\infty \ge C_0\sqrt{\frac{t}{n}}) \le 2p \exp{(-t)}.
\end{equation*}
\QEDB

 {\bf Proof of Theorem \ref{Tm:1}}\\
 By letting $t=(1+C_2)\log p$ in Lemma \ref{Lemma: gradient_bound}, we have
 \begin{equation*}
     P(\|\nabla\mL_{n}(\bb^*)\|_\infty \le C_1\sqrt{\frac{\log p}{n}}) \le 1- 2\exp{(-C_2\log p)}
 \end{equation*}
 as desired for $n \ge (1+C_2) \log p$, where $C_1=C_0\sqrt{(1+C_2)}$. Next we provide the proof of Theorem \ref{Tm:1} (ii). We first suppose the existence of stationary points in the local RSC region and will establish this fact at the end of the proof. Suppose $\hat{\bb}$ is a stationary point of program (\ref{eq:ex-group-penalization}) such that $\|\hat{\bb} - \bb^*\|_2 \le r$. Since
$\hat{\bb}$ is a stationary point and $\hat{\bb}$ is feasible, we have the inequality
\begin{equation}\label{ineq: stat}
    \langle \nabla \mL_n(\hat{\bb}) - \nabla q_\lambda(\hat{\bb}) + \lambda \bD \Tilde{\bz}, \bb^*-\hat{\bb} \rangle \ge \bf 0,
\end{equation}
where $\bD := \text{diag}((\sqrt{d_1}\bone_{d_1}^T,\cdots, \sqrt{d_J}\bone_{d_J}^T)^T)$ denotes a $p \times p$ diagonal matrix, $\tilde{\bz}=(\tilde{\bz}_1^T, \cdots, \tilde{\bz}_J^T)^T$ and $\tilde{\bz}_j \in \partial \|\hat{\bb}_j\|_2$. Recall

\begin{equation*}
    \partial\|\hat{\bb}_j\|_2=\begin{cases}
\frac{\hat{\bb}_j}{\|\hat{\bb}_j\|_2} & \text{if } \|\hat{\bb}_j\|_2 \ne 0, \\
\{ \bz:\|\bz\|_2 \le 1, \bz \in \RR^{d_j}\}  & \text{if } \|\hat{\bb}_j\|_2 = 0,
\end{cases}
\end{equation*}
for $j=1,2, \cdots, J$. By the convexity of $\frac{\mu}{2}\|\bb\|_2^2 - q_\lambda(\bb)$, we have
\begin{equation}
    \langle \nabla q_\lambda(\hat{\bb}), \bb^* - \hat{\bb} \rangle \ge q_{\lambda}(\bb^*) - q_\lambda(\hat{\bb}) - \frac{\mu}{2}\|\hat{\bb} - \bb^*\|_2^2.
\end{equation}
So together with inequality (\ref{ineq: stat}) we obtain
\begin{equation*}
    \langle \nabla \mL_n(\hat{\bb}) + \lambda \bD \Tilde{\bz}, \bb^* - \hat{\bb} \rangle \ge q_{\lambda}(\bb^*) - q_\lambda(\hat{\bb}) - \frac{\mu}{2}\|\hat{\bb} - \bb^*\|_2^2.
\end{equation*}
Since $\langle \lambda \bD \Tilde{\bz}, \bb^* - \hat{\bb} \rangle \le \sum_{j=1}^{J} \sqrt{d_j}\lambda\|\bb_j^*\|_2 - \sum_{j=1}^{J} \sqrt{d_j}\lambda\|\hat{\bb_j}\|_2$, this means
\begin{equation} \label{ineq:penalty}
    \langle \nabla \mL_n(\hat{\bb}), \bb^* - \hat{\bb} \rangle \ge \rho_\lambda(\hat{\bb}) - \rho_\lambda(\bb^*) - \frac{\mu}{2} \|\hat{\bb} - \bb^*\|_2^2.
\end{equation}
Let $ \Tilde{ \bnu} := \hat{\bb} - \bb^*$. From the RSC inequality (\ref{eq: RSC}), we have
\begin{equation} \label{ineq:RSC2}
    \langle \nabla \mL_n(\hat{\bb})- \nabla \mL_n(\bb^*), \hat{\bb}-\bb^* \rangle \ge \gamma \|\Tilde{\bnu}\|_2^2 - \tau \frac{\log p}{n}\|\Tilde{\bnu}\|_1^2.
\end{equation}
Combining inequality (\ref{ineq:RSC2}) with inequality (\ref{ineq:penalty}), we then have
\begin{equation}
    (\gamma - \frac{\mu}{2}) \|\Tilde{\bnu}\|_2^2 - \tau \frac{\log p}{n}\|\Tilde{\bnu}\|_1^2 + (\rho_\lambda(\hat{\bb}) - \rho_\lambda(\bb^*)) \le \langle \nabla \mL_n(\bb^*), \bb^* - \hat{\bb} \rangle.
    \end{equation}
So by Holder's inequality, we conclude that
\begin{equation}
    (\gamma - \frac{\mu}{2}) \|\Tilde{\bnu}\|_2^2 - \tau \frac{\log p}{n}\|\Tilde{\bnu}\|_1^2 + (\rho_\lambda(\hat{\bb}) - \rho_\lambda(\bb^*)) \le \|\nabla \mL_n(\bb^*)\|_{\infty}\|\Tilde{\bnu}\|_1.
    \end{equation}
Assume $\lambda \ge 4\|\nabla \mL_n(\bb^*)\|_{\infty}$ and $\lambda \ge 8\tau R \frac{\log p}{n}$, we have
\begin{equation*}
    \begin{split}
        (\gamma - \frac{\mu}{2})\|\Tilde{\bnu}\|_2^2 &\le (\rho_\lambda(\bb^*) - \rho_\lambda(\hat{\bb})) + (2R\tau \frac{\log p}{n} + \|\nabla \mL_n(\bb^*)\|_{\infty})\|\Tilde{\bnu}\|_1\\
        & \le (\rho_\lambda(\bb^*) - \rho_\lambda(\hat{\bb})) + \sum_{j=1}^{J} \sqrt{d_j} (2R\tau \frac{\log p}{n} + \|\nabla \mL_n(\bb^*)\|_{\infty})\|\Tilde{\bnu}_j\|_2\\
        & \le (\rho_\lambda(\bb^*) - \rho_\lambda(\hat{\bb})) + \frac{1}{2}\sum_{j=1}^{J}\sqrt{d_j}\lambda \|\Tilde{\bnu}_j\|_2\\
        & \le (\rho_\lambda(\bb^*) - \rho_\lambda(\hat{\bb})) + \frac{1}{2}(\rho_\lambda(\Tilde{\bnu}) + \frac{\mu}{2}\|\Tilde{\bnu}\|_2^2),
        \end{split}
\end{equation*}
implying that 
\begin{equation} \label{ineq:med}
    0 \le (\gamma - \frac{3\mu}{4})\|\Tilde{\bnu}\|_2^2 \le \rho_\lambda(\bb^*) - \rho_\lambda(\hat{\bb}) +  \frac{1}{2}\rho_\lambda(\Tilde{\bnu}).
    \end{equation}
    Recall $S \subseteq \{1, \cdots, J\}$ includes all indexes of important groups and $|S|=s$. By the assumption \ref{as:penalty} for $\rho$, we have
    \begin{equation*}
       \rho_\lambda(\Tilde{\bnu}_S) = \rho_\lambda(\bb^* - \hat{\bb}_S) \ge \rho_\lambda(\bb^*) - \rho_\lambda(\hat{\bb}_S),
    \end{equation*}
    where $\hat{\bb}_S$ denotes the zero-padded vector in $\RR^{p}$ with zeros on groups in $S^c$. Then starting from inequality (\ref{ineq:med}), we have
    \begin{equation} \label{ineq:med2}
        \begin{split}
            0 &\le (\gamma - \frac{3\mu}{4})\|\Tilde{\bnu}\|_2^2 \\
            & \le \rho_\lambda(\bb^*) - \rho_\lambda(\hat{\bb}) +  \frac{1}{2}\rho_\lambda(\Tilde{\bnu})\\
            & = \rho_\lambda(\bb^*) - \rho_\lambda(\hat{\bb}_S) - \rho_\lambda(\hat{\bb}_{S^c}) + \frac{1}{2}\rho_\lambda(\Tilde{\bnu})\\
            & \le \rho_\lambda(\Tilde{\bnu}_{S}) - \rho_\lambda(\hat{\bb}_{S^c}) + \frac{1}{2}\rho_\lambda(\Tilde{\bnu})\\
            & = \frac{3}{2}\rho_\lambda(\Tilde{\bnu}_{S}) - \rho_\lambda(\Tilde{\bnu}_{S^c}) + \frac{1}{2}\rho_\lambda(\Tilde{\bnu}_{S^c})\\
            & = \frac{3}{2}\rho_\lambda(\Tilde{\bnu}_{S}) - \frac{1}{2}\rho_\lambda(\Tilde{\bnu}_{S^c}).
        \end{split}
    \end{equation}
    
Let $A$ denote the group index set of the first s groups of $\tilde{\bnu}$ with largest $\ell_2$ norm. Recall $d_a = \max_{1 \le j \le J}d_j$, $d_b = \min_{1 \le j \le J}d_j$, $d=\sqrt{\frac{d_a}{d_b}}$. By assumption \ref{as:penalty}(i) and (iv) we have
\begin{equation} \label{ineq:cone bound}
    \begin{split}
       0 \le 3\rho_\lambda(\Tilde{\bnu}_{S}) - \rho_\lambda(\Tilde{\bnu}_{S^c}) 
        &\le 3\sum_{j \in S} \rho(\|\Tilde{\bnu}_j\|_2, \sqrt{d_a} \lambda) - \sum_{j \in S^c} \rho(\|\Tilde{\bnu}_j\|_2, \sqrt{d_b} \lambda)\\
        & \le 3\sum_{j \in A} \rho(\|\Tilde{\bnu}_j\|_2, \sqrt{d_a} \lambda) - \sum_{j \in A^c} \rho(\|\Tilde{\bnu}_j\|_2, \sqrt{d_b} \lambda).\\
    \end{split}
\end{equation}
Let $c:= \max_{j \in A^c} \|\Tilde{\bnu}_j\|_2$ and define $f(t,\lambda):=\frac{t\lambda}{\rho(t,\lambda)}$ for $t,\lambda>0$. By assumption on $\rho$, for any fixed $\lambda \in \RR^+$, function $t \mapsto f(t,\lambda)$ is non-decreasing on $\RR^+$. Thus
\begin{equation} \label{ineq:maxbound1}
\begin{split}
    \sum_{j \in A} \rho(\|\Tilde{\bnu}_j\|_2, \sqrt{d_a} \lambda)\cdot f(c,\sqrt{d_a} \lambda)  & \le \sum_{j \in A} \rho(\|\Tilde{\bnu}_j\|_2, \sqrt{d_a} \lambda) \cdot f(\|\Tilde{\bnu_{j}}\|_2,\sqrt{d_a} \lambda)\\
    & \le \sum_{j \in A} \sqrt{d_a}\lambda \|\Tilde{\bnu}_j\|_2.
\end{split}
\end{equation}
Similarly we also obtain
\begin{equation} \label{ineq:maxbound2}
    \begin{split}
    \sum_{j \in A^c} \rho(\|\Tilde{\bnu}_j\|_2, \sqrt{d_b} \lambda)\cdot f(c,\sqrt{d_b} \lambda)  & \ge \sum_{j \in A^c} \rho(\|\Tilde{\bnu}_j\|_2, \sqrt{d_b} \lambda) \cdot f(\|\Tilde{\bnu_{j}}\|_2,\sqrt{d_b} \lambda)\\
    & \ge \sum_{j \in A^c} \sqrt{d_b}\lambda \|\Tilde{\bnu}_j\|_2.
\end{split}
\end{equation}
Combining inequality (\ref{ineq:cone bound}) with (\ref{ineq:maxbound1}) and (\ref{ineq:maxbound2}) we have
\begin{equation} \label{ineq:med3}
    \begin{split}
         0 &\le 3\rho_\lambda(\Tilde{\bnu}_{S}) - \rho_\lambda(\Tilde{\bnu}_{S^c})  \\
         & \le \frac{1}{ f(c,\sqrt{d_a} \lambda)}( 3\sum_{j \in A} \sqrt{d_a}\lambda \|\Tilde{\bnu}_j\|_2 - \frac{f(c,\sqrt{d_a} \lambda)}{f(c,\sqrt{d_b} \lambda)}\sum_{j \in A^c} \sqrt{d_b}\lambda \|\Tilde{\bnu}_j\|_2)\\
           & \le 3\sum_{j \in A} \sqrt{d_a}\lambda \|\Tilde{\bnu}_j\|_2 - \frac{f(c,\sqrt{d_a} \lambda)}{f(c,\sqrt{d_b} \lambda)}\sum_{j \in A^c} \sqrt{d_b}\lambda \|\Tilde{\bnu}_j\|_2\\
         & = \sqrt{d_a}\lambda(3\sum_{j \in A} \|\Tilde{\bnu}_j\|_2 - \frac{\rho(c, \sqrt{d_b}\lambda)}{\rho(c, \sqrt{d_a}\lambda)} \sum_{j \in A^c} \|\Tilde{\bnu}_j\|_2)\\
         &  \le \sqrt{d_a}\lambda(3\sum_{j \in A} \|\Tilde{\bnu}_j\|_2 - g(d)^{-1} \sum_{j \in A^c} \|\Tilde{\bnu}_j\|_2),
         \end{split}
\end{equation}
where the third inequality follows from
\begin{equation*}
    f(c,\sqrt{d_a} \lambda) \ge \lim_{r \rightarrow 0^+} f(r,\sqrt{d_a} \lambda) = \lim_{r \rightarrow 0^+} \frac{(r-0)\sqrt{d_a} \lambda}{\rho(r,\sqrt{d_a} \lambda) - \rho(0,\sqrt{d_a} \lambda)}=1,
\end{equation*}
and the last inequality follows from assumption \ref{as:penalty}(ii). Hence,
\begin{equation*}
    3 g(d)\sum_{j \in A} \|\Tilde{\bnu}_j\|_2 \ge  \sum_{j \in A^c} \|\Tilde{\bnu}_j\|_2,
\end{equation*}
implying that
\begin{equation} \label{ineq:l1l2}
\begin{split}
    \|\Tilde{\bnu}\|_1 & \le \sum_{j \in A} \|\Tilde{\bnu}_j\|_1 +  \sum_{j \in A^c} \|\Tilde{\bnu}_j\|_1\\
    & \le \sum_{j \in A} \sqrt{d_a}\|\Tilde{\bnu}_j\|_2 +  \sum_{j \in A^c} \sqrt{d_a}\|\Tilde{\bnu}_j\|_2\\
    & \le \sqrt{d_a}(1+3g(d))\sum_{j \in A} \|\Tilde{\bnu}_j\|_2\\
    & \le \sqrt{d_as}(1+3g(d))\|\Tilde{\bnu}\|_2.
\end{split}
\end{equation}
Combing inequalities (\ref{ineq:med2}) and (\ref{ineq:med3}) then gives
\begin{equation*}
    (\gamma - \frac{3\mu}{4})\|\Tilde{\bnu}\|_2^2 \le \frac{1}{2}\sqrt{d_a}\lambda(3\sum_{j \in A} \|\Tilde{\bnu}_j\|_2 - g(d)^{-1} \sum_{j \in A^c} \|\Tilde{\bnu}_j\|_2) \le \frac{3}{2}\sqrt{d_a}\lambda \sum_{j \in A} \|\Tilde{\bnu}_j\|_2 \le
    \frac{3}{2}\sqrt{d_as}\lambda\|\Tilde{\bnu}\|_2,
\end{equation*}
from which we conclude that
\begin{equation}
    \|\Tilde{\bnu}\|_2 \le \frac{6\sqrt{d_a}\lambda\sqrt{s}}{4\gamma - 3\mu}
\end{equation}
as wanted. Combining the $\ell_2$-bound with inequality (\ref{ineq:l1l2}) yields the $\ell_1$ bound
\begin{equation}
    \|\Tilde{\bnu}\|_1 \le \frac{6(1+3g(d))d_a\lambda s}{4\gamma - 3\mu}.
\end{equation}
Finally, in order to establish the existence of local stationary points, we simply define $\Hat{\bb} \in \RR^p$ such that 
\begin{equation} \label{prgram2}
    \Hat{\bb} \in \argmin_{\|\bb - \bb^*\|_2 \le r, \|\bb\|_1 <R}  \left\{ \mL_n(\bb)+\rho_\lambda(\bb) \right\}.
\end{equation}
Then $\Hat{\bb}$ is a stationary point of program (\ref{prgram2}). Therefore, we have
\begin{equation*}
    \|\Hat{\bb} - \bb^*\|_2 \le C\sqrt{\frac{d_as\log p}{n}}.
\end{equation*}
Provided $n > Cr^{-2}d_as \log p$, the point $\Hat{\bb}$ will lie in the interior of the sphere of radius $r$ around
$\bb^*$. Hence, $\Hat{\bb}$ is also a stationary point of the original program (\ref{eq:ex-group-penalization}), guaranteeing the existence of such local stationary points. \QEDB
\\

To prove Theorem \ref{Tm:2}, we need the following result adopted directly from the Lemma 1 in \cite{loh2017statistical}. 
\begin{lemma} \label{lemma:restrict-cov}
Suppose $\mL_{ n}$ satisfies the local RSC condition (\ref{as:RSC}) and $n \ge \frac{2\tau}{\gamma} k \log p$. Then $\mL_{ n}$ is strongly convex over the region $S_r := \{ \bb \in \RR^p: supp(\bb) \subseteq I_S, \|\bb - \bb^*\|_2 \le r\}$.
\end{lemma}
Proof. The proof is similar to the proof of Lemma 1 in \cite{loh2017statistical}. \QEDB\\

{\bf Proof of Theorem \ref{Tm:2}}\\
The proof is an adaptation of the arguments of Theorem 2 in the paper \cite{loh2017statistical}. We use the following three steps of the primal-dual witness (PDW) construction:
\begin{itemize}
    \item[(i)] Optimize the restricted program 
    \begin{equation} \label{eq:restircted}
        \hat{\bb}_{I_S} \in \argmin_{\bb \in \RR^{I_S}: \|\bb\|_1 \le R}  \left\{  \mL_{ n}(\bb) + \sum_{j \in S} \rho(\|\bb_j\|_2, \sqrt{d_j}\lambda) 
        \right \},
    \end{equation}
    and establish that $\|\hat{\bb}_{I_S}\|_1 < R$.
    
    \item[(ii)] Recall $q_\lambda(\bb) = \sum_{j=1}^{J}\sqrt{d_j}\lambda\|\bb_j\|_2 -  \sum_{j=1}^J \rho(\|\bb_j\|_2\sqrt{d_j}\lambda)$  defined in Section \ref{sec:2}. 
    Define
    $\hat{\bz}_j \in \partial \|\hat{\bb}_j\|_2$ and let 
    $\hat{\bz}_{I_S}=(\hat{\bz}_j^T, j \in S)^T$, and choose $\hat{ \bz}= (\hat{\bz}_{I_S}^T, \hat{\bz}_{I_S^c}^T)^T$ to satisfy the zero-subgradient condition
     \begin{equation} \label{eq:zero-subgradient}
        \nabla \mL_{ n}(\hat{\bb}) - \nabla q_{\lambda}(\hat{\bb})+\lambda \bD\hat{\bz} = \b0,
    \end{equation}
    where $\hat{\bb} := (\hat{\bb}_{I_S}, \pmb 0_{I_S^c})$ and  $\bD = \text{diag}((\sqrt{d_1}\bone_{d_1}^T,\cdots, \sqrt{d_J}\bone_{d_J}^T)^T)$. Show that $\hat{\bb}_{I_S} = \hat{\bb}^\bo_{I_S}$ and establish strict dual feasibility:
    $\max_{j \in S^c}\|\hat{\bz}_j\|_2 < 1$.

    \item[(iii)] Verify via second order conditions that $\hat{\bb}$ is a local minimum of program (\ref{eq:ex-group-penalization}) and conclude that all stationary points $\hat{\bb}$ satisfying $\|\hat{\bb} - \bb^*\|_2 \le r$ are supported on $I_S$ and agree with $\hat{\bb}^\bo$.
\end{itemize}

\textbf{Proof of Step (i) }:  By applying Theorem \ref{Tm:1} to the restricted program (\ref{eq:restircted}), we have
\begin{equation*}
    \|\hat{\bb}_{I_S}-\bb^*_{I_S}\|_1 \le\frac{6(1+3g(d))d_a\lambda s}{4\gamma - 3\mu},
\end{equation*}
and thus
\begin{equation*}
    \|\hat{\bb}_{I_S}\|_1 \le \|\bb^*\|_1 + \|\hat{\bb}_{I_S}-\bb^*_{I_S}\|_1 \le \frac{R}{2} +   \|\hat{\bb}_{I_S}-\bb^*_{I_S}\|_1 \le \frac{R}{2}+\frac{6(1+3g(d))d_a\lambda s}{4\gamma - 3\mu} < R,
\end{equation*}
under the assumption of the theorem. This complete step (i) of the PDW construction. \QEDB
\\

To prove step (ii), we need the following Lemma \ref{Lemma: restrict-bound} and \ref{Lemma: restricted-strict-convex}:
\begin{lemma}  \label{Lemma: restrict-bound}
Under the conditions of Theorem \ref{Tm:2}, we have the bound 
\begin{equation*}
     \|\hat{\bb}^\bo_{I_S}-\bb^*_{I_S}\|_2 \le C_5 \sqrt \frac{\log p}{kn}
\end{equation*}
and $\hat{\bb}_{I_S} = \hat{\bb}^\bo_{I_S}$ with probability at least $1-2 \exp(-C_4 \log p/k^2)$.
\end{lemma}
\textit{Proof.} 
Recall $\hat{\bb}^\bo = (\hat{\bb}^\bo_{I_S}, \b0_{I_S^c})$. By the optimality of the oracle estimator, we have
\begin{equation} \label{ineq:oracle}
    \mL_{ n} (\hat{\bb}^\bo) \le \mL_{ n}(\bb^*).
\end{equation}
Consider $n \ge \frac{2\tau}{\gamma}k\log p$. By Lemma \ref{lemma:restrict-cov} $\mL_{ n}(\bb)$ is strongly convex over restricted region $S_r$. Hence,
\begin{equation}
    \mL_{ n}(\bb^*) + \langle \nabla \mL_{ n}(\bb^*), \hat{\bb}^\bo - \bb^* \rangle + \frac{\gamma}{4} \|\hat{\bb}^\bo - \bb^*\|_2^2 \le \mL_{ n}(\hat{\bb}^\bo).
\end{equation}
Together with inequality (\ref{ineq:oracle}) we obtain
\begin{equation*}
\begin{array}{ll}
\frac{\gamma}{4}  \|\hat{\bb}^\bo - \bb^*\|_2^2 &\le \langle \nabla \mL_{ n}(\bb^*), \bb^* - \hat{\bb}^\bo \rangle \le \| \nabla (\mL_{ n}(\bb^*))_{I_S}\|_\infty \cdot \|\hat{\bb}^\bo - \bb^*\|_1\\
& \le \sqrt{k} \| \nabla (\mL_{ n}(\bb^*))_{I_S}\|_\infty \cdot \|\hat{\bb}^\bo - \bb^*\|_2, 
\end{array}
\end{equation*}
implying that 
\begin{equation} \label{ineq:res-l2-bound}
    \|\hat{\bb}^\bo - \bb^*\|_2 \le \frac{4\sqrt{k}}{\gamma}\| \nabla (\mL_{ n}(\bb^*))_{I_S}\|_\infty.
\end{equation}
By applying Lemma \ref{Lemma: gradient_bound} to the restricted program (\ref{eq:restircted}), we have
\begin{equation*}
   P(\|\nabla (\mL_{ n}(\bb^*_{I_S}))\|_\infty \le C_0\sqrt{\frac{t}{n}}) \ge 1-2k\exp(-t).
\end{equation*}
Let $t=C_3\log p / k^2$. Then we obtain 
\begin{equation}  \label{ineq:res-tm1}
      P(\|\nabla (\mL_{ n}(\bb^*_{I_S}))\|_\infty \le C_0\sqrt{C_3}\sqrt{\frac{\log p}{k^2n}}) \ge 1-2\exp(-C_4 \log p/k^2),
\end{equation}
where we require $k^2\log k=\bo(\log p)$. Combining inequality (\ref{ineq:res-l2-bound}) and (\ref{ineq:res-tm1}), we obtain 
\begin{equation} \label{ineq: res-inf-bound}
     \|\hat{\bb}^\bo - \bb^*\|_2 \le  C_5\sqrt{\frac{\log p}{kn}}
\end{equation}
as desired, where $C_5=4C_0\sqrt{C_3}/\gamma$. 

Next we show $\hat{\bb}_{I_S} = \hat{\bb}^\bo_{I_S}$. When $n > C_5^2/r^2\log p/k$, we have $\|\hat{\bb}^\bo_{I_S} - \bb^*_{I_S}\|_2 < r$ and thus $\hat{\bb}^\bo_{I_S}$ is an interior point of the oracle program in (\ref{eq:oracle}), implying
\begin{equation} \label{eq:oracle-gradient}
    \nabla \mL_{ n}(\hat{\bb}^\bo_{I_S}) = \b0.
\end{equation}
By assumption we have $\lambda=C_6\sqrt{\frac{\log p}{n}}$  and
$\bb^{*G}_{\min} \ge C_8\sqrt{\frac{d_a\log p}{n}}$, where we choose $C_8=C_6\delta + C_5$. Together with inequality (\ref{ineq: res-inf-bound}), we have
\begin{equation*}
    \begin{array}{ll}
    \|\hat{\bb}^\bo_j\|_2 \ge \|\bb^*_j\|_2 - \|\hat{\bb}^\bo_j - \bb^*_j\|_2 &\ge \bb^{*G}_{\min} - \|\hat{\bb}^\bo-\bb^*\|_2 \\
    & \ge ( C_6\delta + C_5)\sqrt{\frac{d_a\log p}{n}} -C_5\sqrt{\frac{\log p}{kn}} \\
    & \ge \sqrt{d_a}\delta \lambda.
    \end{array}
\end{equation*}
for all $j \in S$. Together with the assumption that $\rho$ is $(\mu, \delta)$-amenable, we have 
\begin{equation} \label{eq:oracle-penalty}
    \nabla q_{\lambda}(\hat{\bb}^\bo_{I_S}) = \lambda \bD_{I_SI_S}\hat{\bz}^\bo_{I_S},
\end{equation}
where $\hat{\bz}^\bo_{I_S}=((\hat{\bz}_j^{\bo})^T, j \in S)^T$ and $\hat{\bz}_j^\bo \in \partial\|\hat{\bb}^\bo_j\|_2$. Combining equation (\ref{eq:oracle-gradient}) and (\ref{eq:oracle-penalty}), we obtain
\begin{equation} \label{eq:oracle-zero-subgradient}
     \nabla \mL_{ n}(\hat{\bb}^\bo_{I_S})  - \nabla q_{\lambda}(\hat{\bb}^\bo_{I_S}) + \lambda \bD_{I_SI_S}\hat{\bz}^\bo_{I_S} = \b0.
\end{equation}
Hence $\hat{\bb}^\bo_{I_S}$ satisfies the zero-subgradient condition for the restricted program (\ref{eq:restircted}).
 By step (i) $\hat{\bb}_{I_S}$ is an interior point of the program (\ref{eq:restircted}), then it must also satisfy the same zero-subgradient condition.  Under the strict convexity in Lemma 4,  the  solution that satisfies the zero-subgradient condition is unique. Thus, we obtain $\hat{\bb}_{I_S} = \hat{\bb}^\bo_{I_S}$. \QEDB
\\

The following lemma guarantees that the program in (\ref{eq:restircted}) is strictly convex:

\begin{lemma} \label{Lemma: restricted-strict-convex}
Suppose $\mL_{ n}$ satisfies the local RSC condition (\ref{as:RSC}) and $\rho$ is $\mu$-amenable with $\gamma > \mu$. Suppose in addition the sample size satisfies $n > \frac{2\tau}{\gamma - \mu}k \log p$, then the restricted program in (\ref{eq:restircted}) is strictly convex.
\end{lemma}
\textit{Proof.} This is almost identical to the proof of Lemma 2 in \cite{loh2017support}. We refer the reader to  the arguments provided in that paper. \QEDB \\

\textbf{Proof of step (ii) }: 
We  rewrite the zero-subgradient condition (\ref{eq:zero-subgradient}) as 
\begin{equation*}
    \left (\nabla \mL_{ n}(\hat{\bb}) - \nabla \mL_{ n}(\bb^*) \right ) + \left (\nabla \mL_{ n}(\bb^*)- \nabla q_{\lambda}(\hat{\bb}) \right )+\lambda \bD\hat{\bz} = \b0.
\end{equation*}
Let $\Hat{Q}$ be a $p \times p$ matrix $\hat{Q}= \int_0^1 \nabla^2 \mL_{n} \blp \bb^* + t(\hat{\bb} - \bb^*) \brp dt$. By the zero-subgradient condition and the fundamental theorem of calculas, we have
\begin{equation*}
    \hat{Q}(\hat{\bb} - \bb^*) + \left ( \nabla \mL_{ n}(\bb^*) - \nabla q_{\lambda}(\hat{\bb}) \right ) + \lambda \bD\hat{\bz}=\b0,
\end{equation*}
And its block form is
\begin{equation} \label{eq: block-zero-subgradient}
    \left[ {\begin{array}{*{20}c}
   \hat{Q}_{I_SI_S} & \hat{Q}_{I_SI_S^c}  \\
   \hat{Q}_{I_S^c I_S} & \hat{Q}_{I_S^c I_S^c}  \\    
 \end{array} } \right]
  \left[ {\begin{array}{*{20}c}
   \hat{\bb}_{I_S} - \bb^*_{I_S} \\
   \b0 \\    
 \end{array} } \right]
 + \left[ {\begin{array}{*{20}c}
   \nabla \mL_{ n}(\bb^*)_{I_S} - \nabla q_{\lambda}(\hat{\bb}_{I_S}) \\
   \nabla \mL_{ n}(\bb^*)_{I_S^c} - \nabla q_{\lambda}(\hat{\bb}_{I_S^c}) \\    
 \end{array} } \right] + \lambda 
 \left[ {\begin{array}{*{20}c}
   \bD_{I_SI_S} & \b0  \\
   \b0 & \bD_{I_S^c I_S^c}  \\    
 \end{array} } \right]
 \left[ {\begin{array}{*{20}c}
   \hat{\bz}_{I_S} \\
   \hat{\bz}_{I_S^c} \\    
 \end{array} } \right] = \b0.
\end{equation}

The selection property implies $\nabla q_{\lambda}(\hat{\bb}_{I_S^c})=\b0$. Plugging this result into equation (\ref{eq: block-zero-subgradient}) and performing some algebra, we conclude that
\begin{equation}
    \bD_{I_S^cI_S^c} \hat{\bz}_{I_S^c} = \frac{1}{\lambda}  \left \{ \hat{Q}_{I_S^cI_S}(\bb^*_{I_S} - \hat{\bb}_{I_S}) - \nabla \mL_{ n}(\bb^*)_{I_S^c} \right \}.
\end{equation}
Therefore,
\begin{equation} \label{ineq:pd-feasible}
\begin{array}{ll}
\max_{j \in S^c}\|\hat{\bz}_j\|_2 & \le \max_{j \in S^c}\sqrt{d_j}\|\hat{\bz_j}\|_\infty \\
& = \|\bD_{I_S^cI_S^c}\hat{\bz}_{I_S^c} \|_\infty\\
&=\frac{1}{\lambda}\|\hat{Q}_{I_S^c I_S}(\hat{\bb}_{I_S} - \bb^*_{I_S})- \nabla \mL_{ n}(\bb^*)_{I_S^c}\|_\infty\\
 & \le \frac{1}{\lambda}\|\hat{Q}_{I_S^c I_S}(\hat{\bb}_{I_S} - \bb^*_{I_S})\|_\infty + \frac{1}{\lambda} \|\nabla \mL_{ n}(\bb^*)_{I_S^c}\|_\infty\\
& \le \frac{1}{\lambda} \left \{\max_{j \in I_S^c} \| \be^T_j  \hat{Q}_{I_S^c I_S}\|_2  \right \} \| (\hat{\bb}_{I_S} - \bb^*_{I_S})\|_2 + \frac{1}{\lambda} \|\nabla \mL_{ n}(\bb^*)_{I_S^c}\|_\infty,
\end{array}
\end{equation}
where $ \be_j$ is a standard unit vector with $j$th element being 1.
Observe that 
\begin{equation*}
\begin{array}{ll} 
 [(\be^T_j\hat{Q}_{I_S^cI_S})_m]^2 & \le [\frac{1}{n}\sum_{i=1}^n  w(\bx_i)\bx_{ij} v(\bx_i) \bx_{im}\int_0^1 l''((y_i - \bx_i^T \bb^* - t(\bx_i\hat{\bb} - \bx_i\bb^*))v(\bx_i))dt]^2\\
 & \le  k_2^2 [\frac{1}{n} \sum_{i=1}^n w(\bx_i)\bx_{ij} \cdot v(\bx_i) \bx_{im} ]^2,\\
\end{array}
\end{equation*}
for all $j \in I_S^c$ and $m \in I_S$, where the last inequality follows from assumption \ref{as:loss}(ii). By conditions of Theorem \ref{Tm:2}, the variables $w(\bx_i)\bx_{ij}$ and $v(\bx_i) \bx_{im}$ are both sub-Gaussian. Using standard concentration results for i.i.d sums of products of sub-Gaussian variables, we have
\begin{equation*}
     P([(e_j^T\hat{Q}_{I_S^cI_S})_m]^2 \le C_3') \ge 1-C_2'\exp(-C_3'n).
\end{equation*}
It then follows from union inequality that 
\begin{equation} \label{ineq:maxQ}
    P( \max_{j \in I_S^c} \| e^T_j  \hat{Q}_{I_S^c I_S}\|_2 \le \sqrt{C_3'k}) \ge 1 - C_2'\exp(-C_3'n + \log(k(p-k))) \ge 1-C_2'\exp(-\frac{C_3'}{2}n),
\end{equation}
where $n \ge \frac{2}{C_3'} \log(k(p-k))$.
By Lemma \ref{Lemma: restrict-bound}  we obtain
\begin{equation} \label{ineq:pd-feasible2}
    \|\hat{\bb}_{I_S} - \bb^*_{I_S}\|_2 \le  C_5\sqrt{\frac{\log p}{kn}}.
\end{equation}
Furthermore, Theorem \ref{Tm:1} gives
\begin{equation} \label{ineq:pd-feasible3}
    \|\nabla \mL_{ n}(\bb^*)_{I_S^c}\|_\infty \le \|\nabla \mL_{ n}(\bb^*))\|_\infty \le C_1\sqrt{\frac{\log p}{n}}.
\end{equation}
Combining inequality (\ref{ineq:pd-feasible}), (\ref{ineq:maxQ}), (\ref{ineq:pd-feasible2}) and (\ref{ineq:pd-feasible3}), we have
\begin{equation*}
    \max_{j \in S^c}\|\hat{\bz}_j\|_2 \le \frac{1}{\lambda}C'_6 \sqrt{\frac{\log p}{n}},
\end{equation*}
with probability at least $1-C_7 \exp(-C_4 \log p/k^2)$, where $C'_6=\sqrt{C_3'}C_5+C_1$. In particular, for $\lambda = C_6\sqrt{\frac{\log p}{n}}$ for some $C_6>C'_6$, we conclude at last that the strict dual feasibility condition $ \max_{j \in S^c}\|\hat{\bz}_j\|_2 < 1$ holds, completing step (ii) of the PDW construction.

\textbf{Step (iii) }: Since the proof for this step is almost identical to the proof in Step (iii) of Theorem 2 in \cite{loh2017statistical}, except for the slightly different notations. We refer the reader to  the arguments provided in that paper. \QEDB\\

{\bf Proof of Theorem \ref{Tm:3}}\\
By the condition that
$\bb^{*I}_{\min} \ge C_5\sqrt{\frac{ \log p}{kn}} + \theta$, we have
\begin{equation} \label{ineq:upper_alpha}
    \begin{array}{ll}
    |\hat{\beta}^\bo_j| \ge |\beta^*_j| - |\hat{\beta}^\bo_j - \beta^*_j| &\ge \bb^{*I}_{\min} - \|\hat{\bb}^\bo_{I_S}-\bb^*_{I_S}\|_\infty \\
    & \ge ( C_5\sqrt{\frac{ \log p}{kn}} + \theta) -C_5\sqrt{\frac{ \log p}{kn}} \\
    & = \theta.
    \end{array}
\end{equation}
for all $j \in I_0$, where the second inequality follows from Lemma \ref{Lemma: restrict-bound}. For $j \in I_S - I_0$,
\begin{equation} \label{ineq: lower_alpha}
     |\hat{\beta}^\bo_j| \le \|\hat{\bb}^\bo_{I_S}-\bb^*_{I_S}\|_\infty  \le C_5\sqrt{\frac{ \log p}{kn}} < \theta,
\end{equation}
where the second inequality follows from Lemma \ref{Lemma: restrict-bound} and the last inequality follows from the condition in Theorem \ref{Tm:3}. Recall $\hat{\bb}^\bo = (\hat{\bb}^\bo_{I_S}, \b0_{I_S^c})$. By Theorem \ref{Tm:2} we have $\hat{\bb} = \hat{\bb}^\bo $ with probability at least $1-C_7 \exp(-C_4 \log p/k^2)$.  Together with  (\ref{ineq:upper_alpha}) and (\ref{ineq: lower_alpha}), we have 
\begin{equation*}
    \hat{\bb}^h(\theta) =\hat{\bb}\cdot I(|\hat{\bb}|\ge \theta)
    =\hat{\bb}^{\mathcal{O}}\cdot I(|\hat{\bb}^{\mathcal{O}}|\ge \theta)= (\hat{\bb}_{I_0}^{\mathcal{O}}, \pmb 0_{I_0^c}),
\end{equation*}
as desired. It then gives the result

\begin{equation*}
    \|\hat{\bb}^h(\theta) - \bb^*\|_2 \le  \|\hat{\bb}^\bo_{I_S}-\bb^*_{I_S}\|_2 \le C_5 \sqrt \frac{\log p}{kn},
\end{equation*}
where the last inequality follows from Lemma \ref{Lemma: restrict-bound}.\QEDB

\bibliographystyle{apalike}      
\bibliography{myref.bib}   
\end{document}